\documentclass[a4paper,fleqn]{cas-dc}

\usepackage[numbers]{natbib}
\setcitestyle{square}
\usepackage{amsmath}  
\usepackage{xcolor}   
\usepackage{listings}
\usepackage{subcaption}
\usepackage[super]{nth}
\usepackage{placeins} 
\usepackage[ruled,vlined,linesnumbered]{algorithm2e} 

\def\tsc#1{\csdef{#1}{\textsc{\lowercase{#1}}\xspace}}
\tsc{WGM}
\tsc{QE}

\usepackage{xcolor,colortbl}
\newcommand{\review}[1]{\textcolor{black}{#1}}
\usepackage{booktabs}
\usepackage[table]{xcolor}
\usepackage{tabularx}
\usepackage{array}

\usepackage{fancyhdr}
\newcommand{\fgcsdoi}{10.1016/j.future.2026.108903}

\fancypagestyle{doifooter}{%
  \fancyhf{}%
  \fancyfoot[L]{\footnotesize R. S. Raigada-Garc\'\i a et al.: Future Generation Computer Systems, DOI: \href{https://doi.org/\fgcsdoi}{\fgcsdoi}}%
  \fancyfoot[R]{\footnotesize Page \thepage}%
}

\newcolumntype{Y}{>{\raggedright\arraybackslash}X}
\newcolumntype{L}[1]{>{\raggedright\arraybackslash}p{#1}}
\newcolumntype{Z}{>{\raggedright\arraybackslash\footnotesize}X}

\makeatletter
\AtBeginDocument{
  \DeclareSymbolFont{fixsymbols}{OMS}{cmsy}{m}{n}
  \DeclareSymbolFont{fixlargesymbols}{OMX}{cmex}{m}{n}
  \let\sum\relax
  \DeclareMathSymbol{\sum}{\mathop}{fixlargesymbols}{"50}
  \let\langle\relax
  \let\rangle\relax
  \DeclareMathDelimiter{\langle}{\mathopen}{fixsymbols}{"68}{fixlargesymbols}{"0A}
  \DeclareMathDelimiter{\rangle}{\mathclose}{fixsymbols}{"69}{fixlargesymbols}{"0B}
}
\makeatother
\newcommand{\bigO}[1]{\mathcal{O}\mbox{(}#1\mbox{)}}

\definecolor{fgcs@header}{RGB}{230,237,245}
\definecolor{fgcs@best}{RGB}{220,240,220}

\begin{document}
\pagestyle{doifooter}
\let\WriteBookmarks\relax
\def\floatpagepagefraction{1}
\def\textpagefraction{.001}

\shorttitle{Wave-based dispatch for circuit cutting in hybrid HPC--quantum systems}
\shortauthors{R. S. Raigada-Garc\'\i a et al.}

\title [mode = title]{Wave-based dispatch for circuit cutting in hybrid HPC--quantum systems}

\author[1,2]{Ricard S. Raigada-Garc\'\i a}[orcid=0009-0009-9684-4745]
\ead{ricard.santaigo.raigada@upc.edu}
\credit{Writing -- original draft, Software, Methodology, Investigation, Formal analysis}

\author[1]{Josep Jorba}[orcid=0000-0002-5810-4748]
\cormark[1]
\ead{jjorbae@uoc.edu}
\credit{Writing -- review \& editing, Validation, Investigation}

\author[3]{Sergio Iserte}[orcid=0000-0003-3654-7924]
\ead{sergio.iserte@bsc.es}
\credit{Writing -- review \& editing, Resources, Investigation, Conceptualization}

\affiliation[1]{organization={Universitat Oberta de Catalunya (UOC)},
                city={Barcelona}, country={Spain}}
\affiliation[2]{organization={Universitat Polit\`ecnica de Catalunya (UPC)},
                city={Barcelona}, country={Spain}}
\affiliation[3]{organization={Barcelona Supercomputing Center (BSC)},
                city={Barcelona}, country={Spain}}

\cortext[cor1]{Corresponding author}


\begin{abstract}
Hybrid High-performance Computing (HPC)---quantum workloads based on circuit cutting decompose large quantum circuits into independent fragments, but existing frameworks tightly couple cutting logic to execution orchestration, preventing HPC centers from applying mature resource management policies to Noisy Intermediate-Scale Quantum (NISQ) workloads. We present \textbf{DQR (Dynamic Queue Router)}, a runtime framework that bridges this gap by treating circuit fragments as first-class schedulable units. The framework introduces a backend-agnostic fragment descriptor to expose structural properties without requiring execution layers to parse quantum code, a wave-based coordinator that achieves pipeline concurrency via non-blocking polling, and a production-ready implementation on the \textbf{CESGA Qmio supercomputer} integrating both QPUs \textbf{local on-premises (Qmio)} and \textbf{remote cloud (IBM Torino)} backends. Experiments on a 32-qubit Hardware-Efficient Ansatz (HEA) circuit \review{demonstrate} not only makespan improvements over a monolithic CPU baseline but also \textbf{transparent per-fragment failover recovery} --- specifically rerouting tasks from the local QPU to classical simulators upon encountering hardware-level incompatibilities --- without pipeline restart. For deeper circuits, the coordination residual accounts for only 5\% of the total execution time, highlighting the framework's scalability. These results show that DQR enables HPC centers to integrate NISQ workloads into existing production infrastructure while preserving the flexibility to adopt improved cutting algorithms or heterogeneous backend technologies.
\end{abstract}


\begin{keywords}
 \sep High-performance Computing
 \sep Quantum Computing
 \sep HPC--QC Convergence
 \sep Circuit Cutting
 \sep Resource Management
\end{keywords}

\maketitle
\thispagestyle{doifooter}

\section{Introduction}\label{sec:intr}
Quantum computing (QC) promises exponential speedups for certain problems in chemistry, materials science, and optimization. However, current and near-term quantum processors are constrained by noise, limited qubit counts, and shallow circuit depths, falling into the Noisy Intermediate-Scale Quantum (NISQ) regime~\cite{Preskill2018NISQ}. In this regime, monolithic quantum circuits often cannot fit on a single device, necessitating decomposition techniques. In this regard, hybrid quantum--classical strategies have emerged to mitigate NISQ limitations: variational quantum algorithms (VQAs) use classical optimizers to sidestep qubit/depth limits~\cite{wang_noise-induced_2021,endo_hybrid_2021}; tensor networks classically approximate high-entanglement states~\cite{berezutskii_tensor_2025}; Trotterization discretizes long evolutions into shallow steps~\cite{fratus_describing_2025}; circuit knitting partitions via classical correlations~\cite{piveteau_circuit_2024}; and circuit cutting decomposes large circuits (via gates or wires) into parallel fragments executed independently and reassembled statistically~\cite{tang_cutqc_2021,mitarai_constructing_2021,peng_simulating_2020}.

Circuit cutting garners special interest for its unbiased estimator of the full circuit's exact expectation value (unlike Trotterization's Trotter error or VQAs' ansatz limits) and its scalability across high-performance computing (HPC) clusters~\cite{tejedorninou2025}, making it pivotal for HPC--QC convergence.

However, circuit cutting alone does not solve the system-level (HPC--QC) problem: partitioned fragments still demand planning, routing, execution, monitoring, and dynamic scheduling across heterogeneous resources --- central processing units (CPUs), graphics processing units (GPUs), quantum processing units (QPUs) --- with varying costs, capacities, latencies, and availability. Thus, efficient workloads require not just quantum compilation but also resource and runtime management. Existing hybrid HPC--QC approaches treat QPUs as coarse external accelerators, relegating classical HPC to pre/post-processing and underusing its distributed/adaptive power. Likewise, cutting frameworks emphasize partitioning/reconstruction but bind execution to static workflows, static backends, or framework-specific control paths, neglecting fine-grained orchestration.

While circuit cutting has advanced algorithmically, a critical systems gap remains: existing frameworks couple cutting logic tightly to execution orchestration, preventing HPC centers from applying mature scheduling, fault tolerance, and heterogeneous routing policies to quantum workloads. Prior work has focused either on algorithmic decomposition~\cite{Elsharkawy2023IntegrationQAcc} or on static frameworks for small-scale experiments~\cite{Beck2024IntegratingQC,Shehata2024HPCQCFramework}, leaving unaddressed the need for a runtime that treats fragmented quantum circuits as first-class citizens in production HPC workloads.

We present a systems framework that bridges this gap by decoupling quantum circuit cutting from execution orchestration in hybrid HPC--QC environments. Our contributions are threefold:

\begin{enumerate}
    \item \textbf{Fragment abstraction}: We introduce a standardized, immutable fragment descriptor tuple exposing structural properties (qubits, depth, gates, reconstruction coefficient, backend admissibility) without requiring execution layers to parse quantum circuit code. This enables backend-agnostic routing and allows cutting frameworks to evolve independently.
    \item \textbf{Wave-based dynamic orchestration}: We design a coordinator-driven, wave-based dispatch algorithm that achieves pipeline concurrency via non-blocking polling, handles transient QC failures through retry/failover mechanisms, and adapts to capacity changes from the resource manager. The runtime supports iteration-aware policies that prioritize scarce QPU slots while maintaining high utilization.
    \item \textbf{Production-ready implementation}: We present an implementation of the complete ecosystem in the supercomputer Qmio at the Galician Supercomputing Center (CESGA), integrating Qdislib for cutting, Qulacs for CPU simulation, and Qmio SDK for NISQ QPU execution. 
\end{enumerate}

Figure~\ref{fig:runtime-architecture} positions each contribution within the architecture it addresses.

Our approach directly addresses gaps highlighted in recent HPC--QC surveys~\cite{Dobler2025SurveyHPCQC}: the lack of standardized fragment descriptors for heterogeneous routing, the absence of decoupled cutting--execution interfaces, and the need for dynamic replanning in the face of noisy, unreliable QPU backends. By treating fragmented circuits as schedulable units with explicit resource requirements, we enable HPC centers to integrate NISQ workloads into existing production workloads while preserving the flexibility to adopt improved cutting algorithms or backend technologies.

The remainder of this paper is organized as follows. 
Section~\ref{sec:back} establishes the technical background on circuit cutting, Qdislib, and hybrid HPC--QC execution.
Section~\ref{sec:rela} discusses related work.
Section~\ref{sec:meth} details our methodology: the execution model, fragment abstraction, and wave-based dispatch algorithm. 
Section~\ref{sec:impl} presents the implementation architecture, backend adapters, and execution flow. Section~\ref{sec:eval} describes experimental evaluation on CESGA's HPC cluster with NISQ QPU access. Section~\ref{sec:disc} discusses key properties and limitations of the
framework. Section~\ref{sec:conc} concludes.

\section{Background}\label{sec:back}
This section establishes the technical context for the Dynamic Queue Router (DQR) framework. We first describe circuit cutting as the decomposition technique that produces the independent fragments DQR manages (\ref{subsec:cutting}), then introduce Qdislib as the cutting library used in this work (\ref{subsec:qdislib}), and finally characterize the hybrid HPC--QC execution problem that arises once circuits are decomposed and must be scheduled across heterogeneous resources (\ref{subsec:hybrid}).

\subsection{Circuit cutting}\label{subsec:cutting}
Circuit cutting overcomes NISQ limitations by decomposing quantum circuits into smaller fragments evaluated independently and recombined via classical post-processing.

\begin{figure}
    \centering
    \includegraphics[clip,width=0.999\linewidth,trim={2.2cm 2.2cm 2.2cm 2.2cm}]{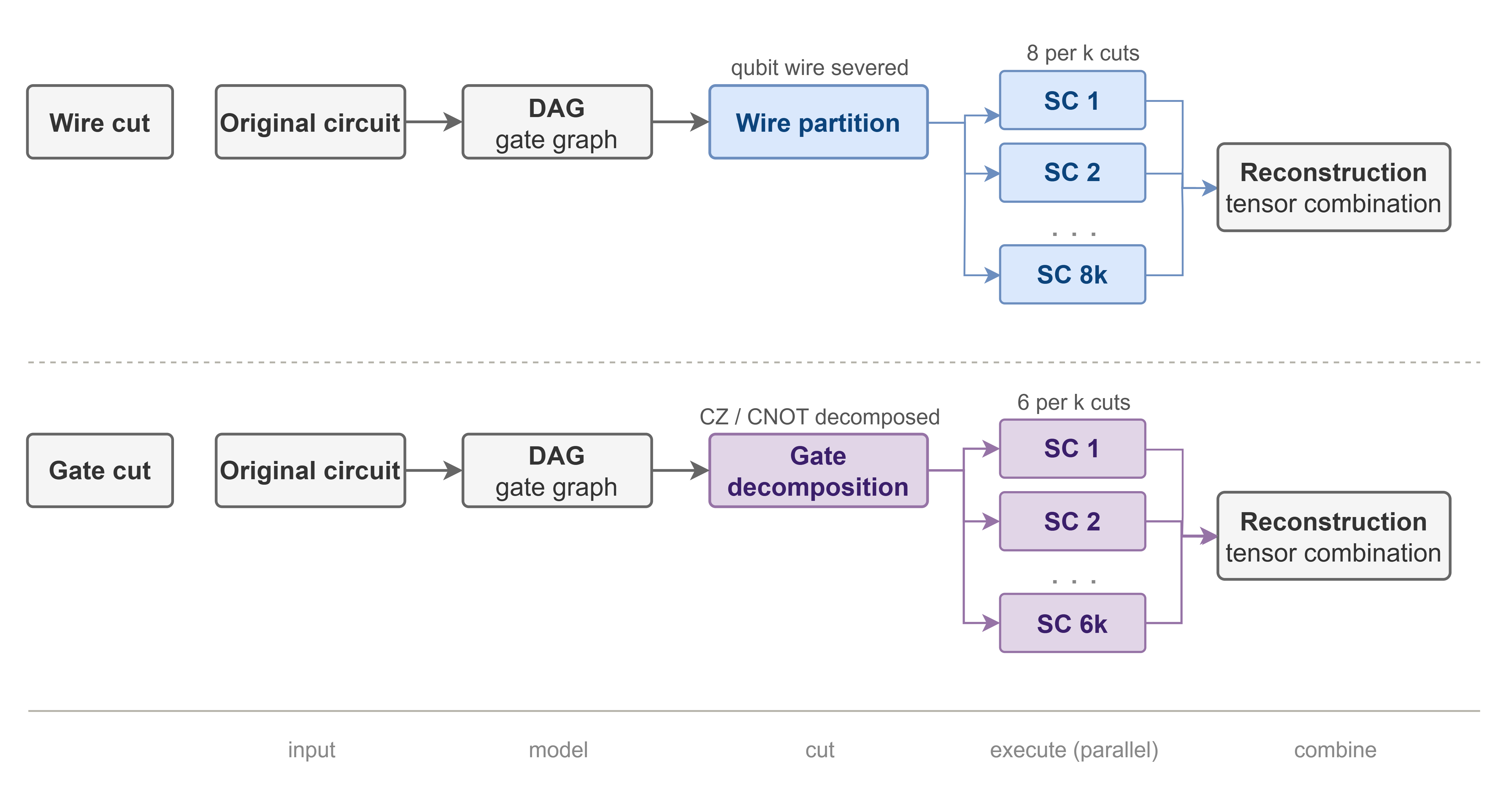}
    \caption{Circuit cutting strategies.
        Both map the circuit to a DAG and split it into independent
        subcircuits (\review{SC}) executed in parallel.
        Wire cutting severs qubit paths ($8^k$ subcircuits), while gate
        cutting decomposes two-qubit gates into quasi-probabilistic
        locals ($6^k$ variants).
        Results are recombined via tensor reconstruction.}
    \label{fig:cutting-example}
\end{figure}

Circuits are modeled as directed acyclic graphs (DAGs): nodes are gates; edges are causal dependencies. This enables framework-agnostic cuts, illustrated in Figure~\ref{fig:cutting-example}: \emph{wire cuts} sever qubit paths (edge between same-qubit gates), yielding up to $8^k$ subcircuits for $k$ cuts; \emph{gate cuts} decompose multi-qubit gates (e.g., CNOT/CZ) into quasi-probabilistic locals ($6^k$ variants).

Post-execution, observables reconstruct via tensor methods:
\begin{equation}
\langle O \rangle = \sum_i c_i \langle O_i \rangle,
\end{equation}
where $c_i$ are quasi-probability coefficients from cuts/projections.

Cutting scales hardware reach (fragments on CPUs/GPUs/QPUs) at exponential classical overhead---$\bigO{6^k}$ subcircuits for gate cuts and $\bigO{8^k}$ for wire cuts---, transforming monolithic programs into distributable workloads needing orchestration.

\review{Circuit cutting extends the effective reach of the available hardware by producing fragments that can be executed independently on CPUs, GPUs, or QPUs, at the cost of an exponential number of subcircuits---$\bigO{6^k}$ for gate cuts and $\bigO{8^k}$ for wire cuts, where $k$ is the number of cuts---which in turn requires dedicated orchestration to remain distributable at scale.}

From a systems view, fragments expose traits (qubits/depth/ops/entanglement) for heterogeneous dispatch, motivating runtime management.

\subsection{Qdislib}\label{subsec:qdislib}

Qdislib\footnote{\url{https://github.com/bsc-wdc/qdislib}} is a Python library developed by the Barcelona Supercomputing Center (BSC) for scalable quantum circuit execution using circuit cutting techniques. It decomposes large quantum circuits into smaller subcircuits for parallel execution on HPC resources, GPUs, or QPUs, overcoming hardware qubit limitations.

The library supports both wire and gate cutting with quasi-probabilistic reconstruction and integrates with quantum frameworks such as Qiskit and Qibo.

It leverages PyCOMPSs for task-based parallelism, enabling hybrid workflows on local simulators (e.g., Qiskit Aer, cuQuantum, Qibojit) or remote QPUs like IBM Quantum~\cite{10.1145/3731599.3767547}.

\subsection{Hybrid HPC--QC execution}\label{subsec:hybrid}

Hybrid HPC--QC execution refers to the integration of classical HPC infrastructures with QPU to cooperatively execute quantum-classical workloads. However, in most current systems, this integration remains coarse-grained. Quantum devices are typically exposed as external accelerators accessed via remote APIs, while classical resources handle circuit preparation, parameter optimization, and post-processing tasks.

In this model, quantum circuit execution is usually treated as a monolithic (distributed or batch) operation delegated to a dedicated backend. Therefore, the role of the classical infrastructure is limited to orchestration and auxiliary computation, rather than actively participating in the execution of the quantum workload. While this approach is sufficient for many hybrid algorithms, it does not fully leverage the distributed capabilities of modern HPC systems.

Circuit cutting changes the execution model. Once a circuit is decomposed, the resulting subcircuits become independent computational fragments that can differ significantly in qubit requirements, circuit depth, and execution cost. Consequently, different fragments may be better suited to different types of computational resources. For example, fragments with few qubits can be efficiently simulated on CPUs or GPUs, while others may benefit from execution on dedicated QPUs.

This transformation introduces a distributed execution problem where fragment scheduling, backend selection, and resource allocation must be performed dynamically across heterogeneous infrastructures. Therefore, efficient execution requires an orchestration layer capable of managing large sets of circuit fragments, considering both workload characteristics and system constraints, including hardware availability, execution capacity, and specific backend limitations.

Despite recent advancements in circuit cutting frameworks and hybrid execution environments, most existing solutions tightly tie circuit decomposition to specific execution workflows or runtime frameworks. This commonly involves transferring fragments from the user side to a queuing system based on hardware type and delegating responsibility to the vendor's queuing system. As a result, circuit fragment scheduling and placement are often static or framework-dependent, limiting the ability to adapt execution strategies to heterogeneous HPC--QC infrastructures.

These limitations lead us to the need for runtime-level approaches that decouple circuit decomposition from execution orchestration and allow for flexible allocation of fragments across classical and quantum resources. In this context, circuit fragments can be treated as independent computational units whose execution can be dynamically routed according to system policies and hardware capabilities.

\section{Related Work}\label{sec:rela}

The integration of quantum computing resources to HPC environments has emerged as a key research topic in the NISQ era. Preskill's seminal work~\cite{Preskill2018NISQ} on NISQ devices argues that near-term quantum advantage is most likely to be achieved via \emph{hybrid} quantum-classical schemes rather than fully fault-tolerant quantum computers. This perspective motivates viewing QPUs as accelerators that complement, rather than replace, classical supercomputers.


One of the first systematic architectural studies of HPC--QC is due to Britt and Humble, who analyze how QPUs can be integrated into current and future HPC system architectures~\cite{Britt2015HPCQPU}. They distinguish between \emph{tight} integration, in which QPUs are attached to compute nodes as accelerators, and \emph{loose} integration, in which QPUs are hosted as remote services accessed over a network. Their work highlights the role of a quantum interconnect in entangling multiple QPUs and argues that conventional performance metrics are insufficient, calling instead for metrics that capture the interplay between system architecture and quantum parallelism~\cite{Britt2015HPCQPU}.

Building on these ideas, subsequent work at large-scale facilities has focused on end-to-end ecosystems rather than isolated devices. Beck \emph{et~al.}~\cite{Beck2024IntegratingQC} propose a hardware-agnostic framework for integrating quantum computing resources---both physical devices and simulators---into scientific HPC workflows, treating quantum resources as accelerators within production environments and demonstrating this in DOE mission applications. Complementing this, Shehata \emph{et~al.}~\cite{Shehata2024HPCQCFramework} define a detailed integration framework and requirements specification (Quantum Framework, QFw) that decomposes an HPC--QC system into resource managers, quantum task managers, platform managers, and runtime controllers, and enumerates usage patterns and integration models relevant to batch-scheduled, MPI-based HPC systems.

Recent frameworks have also explored orchestration of hybrid quantum--classical workflows at higher abstraction levels. The Kubernetes-native framework of Tejedor \emph{et~al.}~\cite{tejedor2026kubernetesorchestratedhybridquantumclassicalworkflows} demonstrates unified management of CPUs, GPUs, and QPUs via Argo Workflows and Kueue, including a proof-of-concept distributed circuit-cutting workflow. Their work bridges the gap between cloud-native orchestration and hybrid quantum pipelines, providing reproducibility, monitoring, and resource-aware scheduling across heterogeneous nodes. However, it operates at the \emph{workflow stage} level rather than exposing individual fragments as schedulable units, lacking the fine-grained fragment-level routing, dynamic backend failover, and policy-driven dispatch that our perspective enables for circuit cutting.


As prototypes and frameworks have proliferated, several surveys have systematized the emerging literature. Döbler and Jattana~\cite{Dobler2025SurveyHPCQC} present a comprehensive survey of works that integrate quantum computers into HPC systems, classifying more than one hundred publications by hardware architecture, software stack, workflow integration, and application domain. They identify a fragmented ecosystem with many point solutions and argue for standardized interfaces and methods to enable interoperability and reuse.

From a software and programming-tools perspective, Elsharkawy \emph{et~al.}~\cite{Elsharkawy2023IntegrationQAcc} review quantum programming tools with a specific focus on their suitability for HPC integration. They introduce a taxonomy based on criteria such as host-language support, execution model, compilation and optimization capabilities, scalability, and support for heterogeneous back-ends. Moreover, they relate tools to different hardware-integration scenarios (standalone quantum systems, co-located accelerators, and on-node integration), thereby providing guidance on selecting or toolchains for HPC--QC deployments.


On the algorithmic side, the state of the art in hybrid quantum-classical algorithms directly shapes the workloads targeted by HPC--QC systems. Endo \emph{et~al.}~\cite{endo_hybrid_2021} survey hybrid algorithms such as the variational quantum eigensolver (VQE), the quantum approximate optimization algorithm (QAOA), and related variational schemes, and they review quantum error-mitigation techniques appropriate for NISQ devices. Their work clarifies the computational structure and classical-quantum interaction patterns (optimization loops, sampling requirements) that HPC architectures must support, while also underscoring that error mitigation can dramatically increase sampling cost and thus classical resource demand.

More broadly, the optimization landscape of hybrid algorithms has been scrutinized in terms of trainability and barren plateaus. Ge \emph{et~al.}~\cite{Ge2022OptimizationLandscape} analyze the optimization landscapes arising in hybrid quantum-classical algorithms, connecting insights from quantum control to NISQ applications and identifying conditions under which gradients vanish or remain robust. These results have important implications for system-level design because they influence the number of circuit evaluations and classical iterations required, and therefore the load imposed on both QPUs and classical HPC resources.


A significant fraction of proposed near-term HPC--QC applications fall into the broad category of quantum-enhanced machine learning. De~Luca~\cite{DeLuca2021SurveyNISQML} surveys hybrid quantum-classical machine learning approaches in the NISQ era, including variational quantum circuits, quantum kernels, and quantum-assisted feature maps. While most demonstrations are still small-scale, the survey highlights the architectural patterns---tight classical--quantum feedback loops and data-intensive pre- and post-processing---that make HPC resources particularly relevant for scaling such workloads.


Dynamic resource allocation has also been explored in recent HPC--QC work. Rocco \emph{et~al.}~\cite{rocco_dynamic_2025} propose both a workflow-based strategy and a malleability-based approach to release classical resources during quantum execution and reallocate them afterward, showing improved resource utilization and time-to-solution over a statically allocated baseline.

A persistent gap in the above ecosystem is the absence of fine-grained, runtime-level orchestration for heterogeneous QPU workloads. McCaskey \emph{et~al.}~\cite{McCaskey2020XACC} introduce XACC, a service-oriented middleware that exposes QPUs as coprocessors to HPC nodes and underpins many subsequent integration stacks. At the language level, QCOR~\cite{Mintz2020QCOR} and its compiler implementation~\cite{Nguyen2022QCOR} extend C++ with quantum kernels compiled against XACC backends, establishing an MPI-compatible programming model for extreme-scale systems. More recently, Mantha \emph{et~al.}~\cite{Mantha2025PilotQuantum} propose Pilot-Quantum, a middleware that applies the Pilot Abstraction from distributed HPC to manage quantum resources and task queues across Slurm-scheduled clusters and cloud QPUs. On the scheduling front, Giortamis \emph{et~al.}~\cite{Giortamis2025Qonductor} demonstrate with Qonductor that jointly optimizing fidelity and job-completion time yields substantial reductions in QPU wait times in cloud settings, while the analyses of Viviani \emph{et~al.}~\cite{Viviani2024Scheduling} argue that scheduling is the primary bottleneck in HPC--QC integration and identify QPU scarcity, technology heterogeneity, and software ecosystem mismatch as root causes. Despite these advances, none of the above systems exposes \emph{individual circuit fragments} as first-class schedulable units: they target whole-circuit jobs or coarse workflow stages, and none provides the combination of a backend-agnostic fragment descriptor, dynamic wave-based dispatch, and policy-driven failover that production circuit-cutting workloads require~\cite{Britt2015HPCQPU, Beck2024IntegratingQC, Shehata2024HPCQCFramework}.

Combining these strands, the state of the art in HPC--QC can be considered a convergence of (i) system architectures that treat QPUs as accelerators within heterogeneous supercomputers, (ii) integration frameworks and surveys that articulate requirements and classify available tools, and (iii) a growing body of hybrid algorithms and applications whose computational structure is well-matched to classical--quantum co-processing in HPC environments.

\section{Methodology}\label{sec:meth}

This section describes the rationale behind the presented system framework that decouples quantum circuit cutting from HPC execution orchestration, enabling scalable hybrid HPC--QC execution of NISQ workloads. The system introduces a standardized fragment abstraction, dynamic wave-based scheduling, and fault-tolerant routing policies that address gaps in prior work~\cite{Britt2015HPCQPU, Beck2024IntegratingQC, Elsharkawy2023IntegrationQAcc, Dobler2025SurveyHPCQC}: lack of cutting-execution decoupling, standardized descriptors for heterogeneous routing, and dynamic re-planning for unreliable NISQ backends.


\begin{figure*}[H]
    \centering
    \includegraphics[clip,width=0.9\textwidth,trim={3.2cm 2.2cm 2.2cm 2.2cm}]{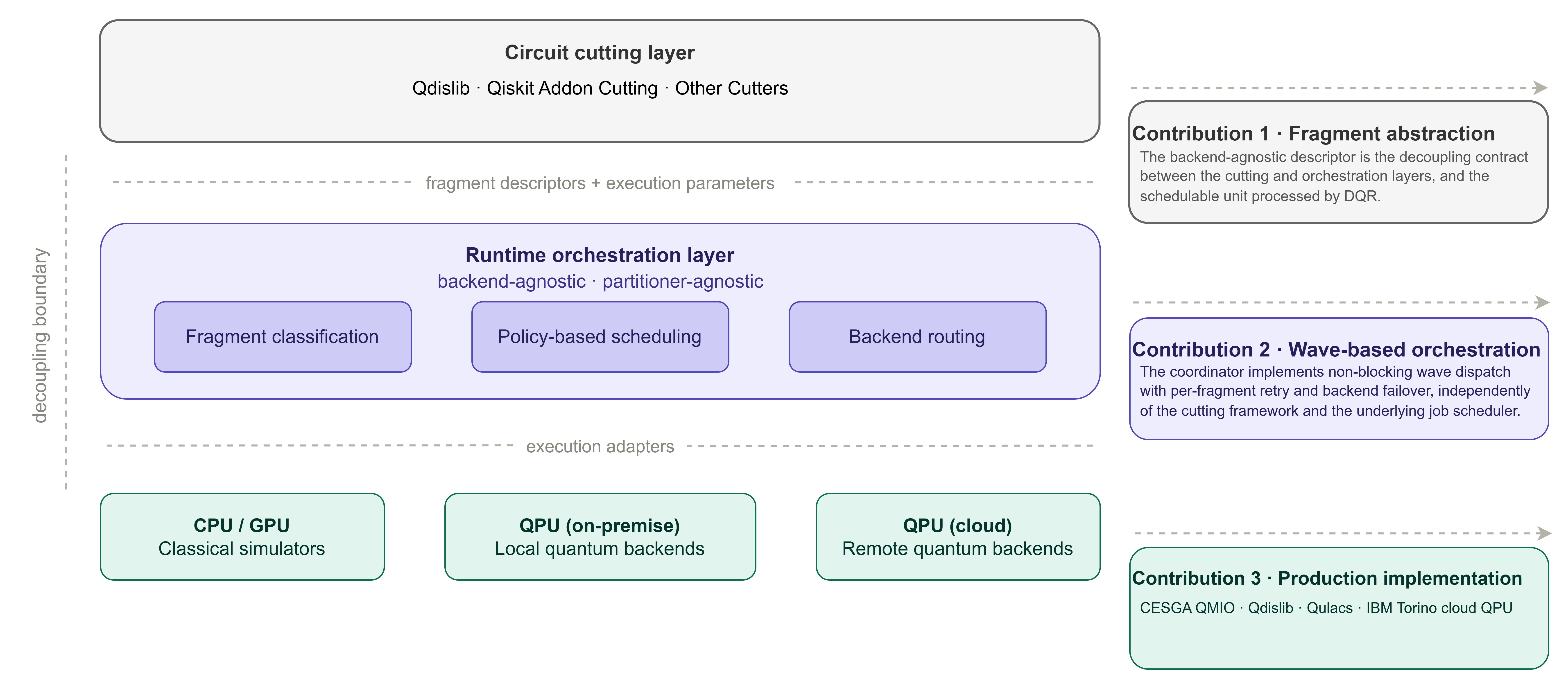}
    \caption{Runtime-oriented execution model for circuit-cut workloads. The circuit cutting layer is decoupled from the orchestration layer via fragment descriptors, enabling hardware-aware, policy-driven scheduling across heterogeneous HPC--QC backends. The three contributions of this work (right) are positioned within the layer they address; see Section~\ref{sec:intr}.}
    \label{fig:runtime-architecture}
\end{figure*}

Figure~\ref{fig:runtime-architecture} showcases the layered approach proposed in this paper. The execution model assumes a pre-cut quantum circuit decomposed into independent fragments via any cutting framework. The core principle is \emph{decoupling of cutting from orchestration}: the cutting layer produces lightweight fragment descriptors exposing only structural properties (qubits, depth, gates, reconstruction coefficient, backend admissibility); the execution layer consumes these descriptors without knowledge of the quantum algorithm, Pauli strings, or cutting method.

This separation transforms a quantum decomposition problem into a classical heterogeneous scheduling problem amenable to HPC techniques: resource allocation, backend routing, fault tolerance, and tensor reconstruction. In the NISQ context---where circuits exceed single-device qubit/depth limits---it enables modularity: cutting frameworks evolve independently while execution scales with HPC resources.

In this regard, a fragment $f_i$ is represented as an immutable tuple:
\begin{equation}
    f_i = \bigl(\mathit{SC}_i,\, q_i,\, d_i,\, g_i,\, c_i,\, \mathcal{B}_i\bigr),
    \label{eq:fragment}
\end{equation}
where:
\begin{itemize}
\item $\mathit{SC}_i$: subcircuit payload (OpenQASM string + metadata),
\item $q_i$: qubit count,
\item $d_i$: circuit depth,
\item $g_i$: two-qubit gate count,
\item $c_i$: quasi-probability coefficient for tensor reconstruction,
\item $\mathcal{B}_i \subseteq \{\text{HPC},\text{QC}\}$: execution backends.
\end{itemize}
While structural metrics $(q_i,d_i,g_i)$ are computed once by cutting, the backend ($\mathcal{B}_i$) where the fragment will be executed is determined during the labeling stage.

When labeling, soft hints (\texttt{HPC}, \texttt{QC}, and \texttt{Undecided}) with the most appropriate backend are assigned to each fragment following a  labeling policy that determines where the fragment will be firstly executed.
In this regard, this presented methodology enables backend-agnostic routing without circuit inspection, since the scheduler does not need to parse or analyze the actual quantum circuit code to decide where to send each fragment.

\begin{figure}
    \centering
    \includegraphics[width=\columnwidth]{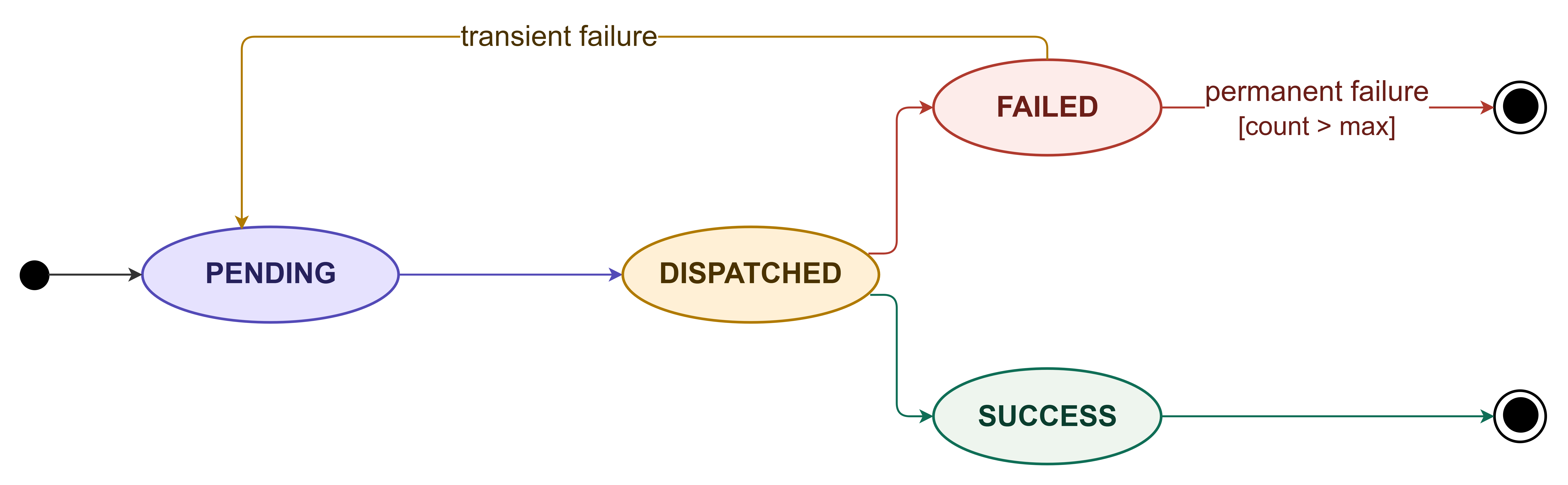}
    \caption{DQR Fragment lifecycle state machine. A fragment is created in the \texttt{PENDING} state and transitions to \texttt{DISPATCHED} upon assignment to a backend. A transient failure returns the fragment to \texttt{PENDING} for retry; once the retry counter exceeds the configured maximum, the fragment reaches the terminal \texttt{PERMANENT\_FAILED} state. Successful execution transitions the fragment to the terminal \texttt{SUCCESS} state, at which point its expectation value is stored for tensor reconstruction.}
    \label{fig:state-machine}
\end{figure}

Execution proceeds in discrete \emph{dispatch waves} via a fragment state machine ($\texttt{PENDING} \to \texttt{DISPATCHED} \to \texttt{SUCCESS}/\texttt{FAILED}$) (see Figure~\ref{fig:state-machine}). 
While the coordinator process non-blocking polls completions from the backends, in each wave it:
\begin{enumerate}
\item Handles failures: 
    \begin{itemize}
        \item transient $\to$ retry (up to max)
        \item permanent $\to$ failover
        \item QC $\to$ HPC if enabled
    \end{itemize}
\item Refreshes capacity (HPC and QC processing slots).
\item Plans dispatch $\Pi$ respecting labels/policies.
\item Submits HPC to free workers, QC to backend session.
\item Commits $\Pi$, incrementing wave counter.
\end{enumerate}

\begin{algorithm}[t]
\DontPrintSemicolon
\SetAlgoLined
\KwIn{Fragments $F$, capacity $C$}
\Repeat{$|\texttt{PENDING}|=|\texttt{DISPATCHED}|=0$}{
  Non-blocking poll completions\;
  \ForEach{completion $c$}{
      \uIf{{$\texttt{TRANSIENT\_FAILED}(c.id)$}}{
        \uIf{retries $<$ max}{
            \uIf{failover}{Re-label $\texttt{QC}\to\texttt{HPC}$}   
            $\texttt{PENDING}(c.id)$}
        \Else{$\texttt{PERMANENT\_FAILED}(c.id)$}
        }
        \Else{$\texttt{SUCCESS}(c.id)$}
    }
  Refresh $C$ (HPC and QC slots)\;
  $\Pi \leftarrow \texttt{plan}(F_\text{PENDING}, C)$\;
  Submit $\Pi_\text{HPC}$ and $\Pi_\text{QC}$, respectively\;
  Commit: selected $\texttt{PENDING}\to\texttt{DISPATCHED}$\;
  Wave counter $++$\;
}
\caption{Wave-based coordination}
\label{alg:wave}
\end{algorithm}

Algorithm~\ref{alg:wave} presents the wave-based dispatch loop.
Each iteration processes three phases: completion handling, capacity refresh, and dispatch planning.

First, the coordinator non-blocking polls fragment completions from HPC and QC backends.
For each completion $c$, the algorithm distinguishes between successful execution and failures.
\texttt{TRANSIENT\_FAILED} fragments are requeued as \texttt{PENDING} if retries remain, with
optional QC$\to$HPC relabeling via the \texttt{failover} flag; exhausted retries trigger
\texttt{PERMANENT\_FAILED} terminal state. Successful fragments transition to \texttt{SUCCESS},
storing their expectation values for tensor reconstruction.

Second, capacity $C$ is refreshed from current HPC allocation and QC status (available QPU session slots). 

Third, the planner $\texttt{plan}(F_\text{PENDING}, C)$ generates a dispatch plan $\Pi$ respecting
fragment labels, routing policies, and available slots. HPC fragments $\Pi_\text{HPC}$ are
assigned to HPC processing units, while QC fragments $\Pi_\text{QC}$ are submitted to QPUs.
Selected fragments commit $\texttt{PENDING}\to\texttt{DISPATCHED}$ in the state machine.

Non-blocking polling enables wave overlap: new fragments dispatch while prior fragments execute,
achieving pipeline concurrency bounded by total MPI ranks and QC slots. The wave counter
supports iteration-aware policies (wave~0 favors labels; later waves prioritize scarce QC slots).


\section{Implementation}\label{sec:impl}
The implementation follows the architecture introduced in Section~\ref{sec:meth} which comprises the three loosely coupled layers and communicates via files on the General Parallel File System (GPFS) and the Message Passing Interface (MPI) paradigm.

\begin{figure}
    \centering
    \includegraphics[clip,width=0.999\linewidth,trim={4.2cm 2.2cm 2.2cm 2.2cm}]{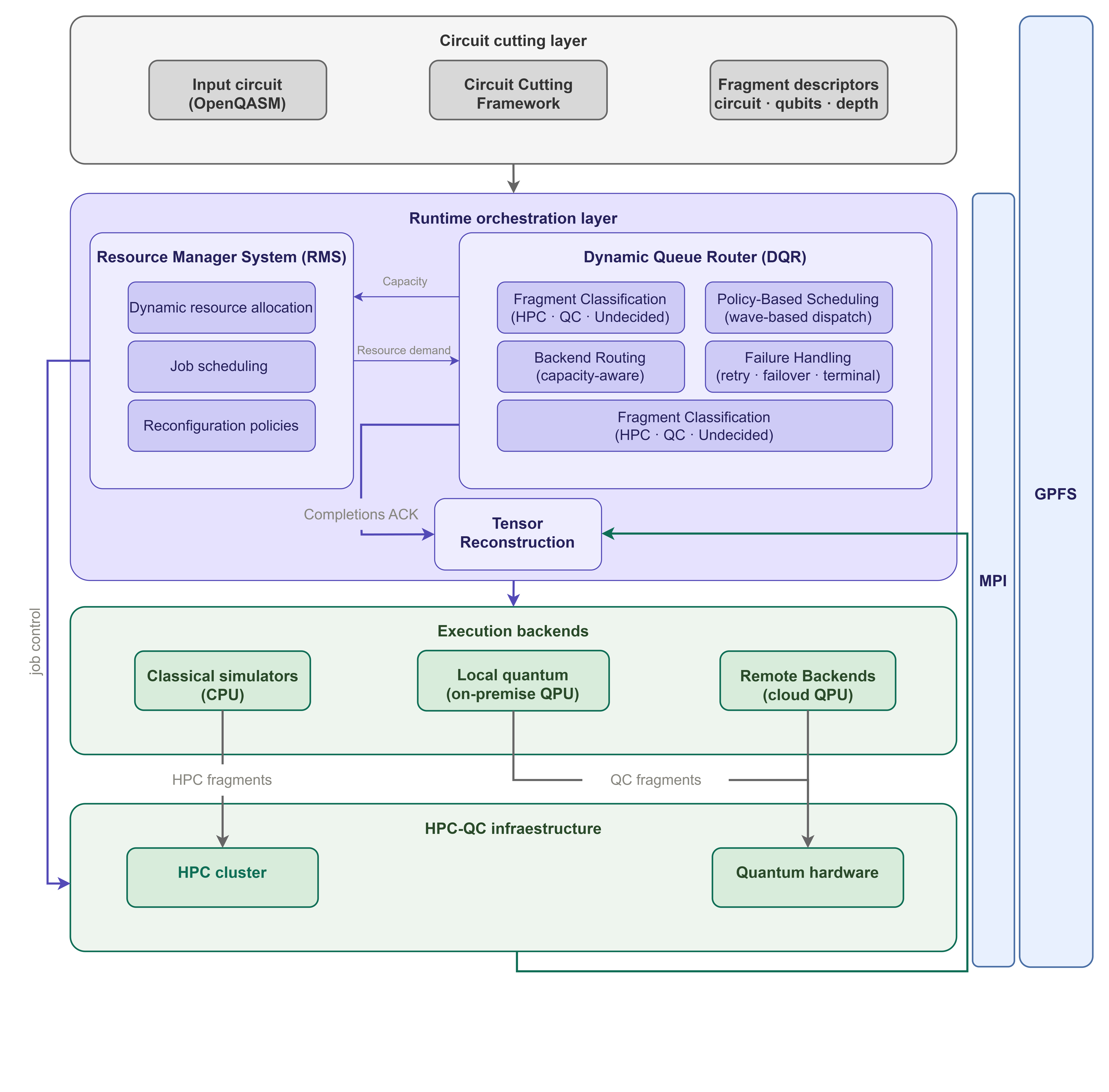}
    \caption{System architecture of the DQR framework. The three loosely coupled layers communicate via GPFS (file-based fragment descriptors and results) and MPI (coordinator--worker messages). The \textit{Circuit Cutting Layer} decomposes the input OpenQASM circuit and emits fragment descriptors. The \textit{Runtime Orchestration Layer} comprises the RMS, which manages HPC allocation and reconfiguration policies, and the DQR, which performs fragment classification, policy-based wave dispatch, capacity-aware backend routing, and failure handling. The \textit{Execution Backends} layer abstracts classical CPU simulators, local on-premises QPUs, and remote cloud QPUs, routing HPC and QC fragments to the corresponding physical infrastructure.}
    \label{fig:dqr-system-architecture}
\end{figure}

Particularly, Figure~\ref{fig:dqr-system-architecture} showcases how the different modules, components, and tools are interconnected.
This section describes in detail the pieces that constitute this system architecture and how they are related.

\subsection{Circuit Cutting Layer}\label{subsec:imp_cut}

The cutting layer is implemented as a gRPC microservice that encapsulates Qdislib. It receives an OpenQASM circuit along with hardware constraints (maximum qubits per subcircuit and max cuts) and returns the complete set of shard descriptors. It is designed so that the Qdislib microservice can be replaced by another cutter as long as the descriptors are respected.


Before entering the dispatch pipeline, each fragment is assigned a placement label by the \texttt{qcut\_label\_job} routine. The labeler reads four structural metrics from each \texttt{.meta.json} file: qubit count~$q_i$, circuit depth~$d_i$, two-qubit gate count~$g_i$, and total operation count~$z_i$. These metrics are mapped to one of three placement labels---\texttt{QC}, \texttt{HPC}, or \texttt{Undecided}---by one of four configurable classification policies. The choice of policy and its thresholds is a \emph{job-preparation} decision made by the operator based on knowledge of the target hardware; the DQR coordinator is entirely label-agnostic and behaves identically regardless of how labels are assigned. What changes with labeling is the prior routing hint, not the correctness or the coordination logic of the system.

\begin{itemize}

\item \emph{Budget.}
A fragment is labeled \texttt{QC} if and only if all available metrics fall within a set of operator-defined upper bounds (the \emph{QC envelope}):
\begin{equation}\label{eq:budget-qc}
  q_i \le q^{\max}_{\mathrm{QC}}
  \;\wedge\;
  d_i \le d^{\max}_{\mathrm{QC}}
  \;\wedge\;
  g_i \le g^{\max}_{\mathrm{QC}}
  \;\wedge\;
  z_i \le z^{\max}_{\mathrm{QC}}.
\end{equation}
A fragment is labeled \texttt{HPC} through a \emph{voting} mechanism: each metric that exceeds its corresponding HPC lower bound ($q_i \ge q^{\min}_{\mathrm{HPC}}$, $d_i \ge d^{\min}_{\mathrm{HPC}}$, etc.) contributes one vote, and the fragment is labeled \texttt{HPC} when the total vote count reaches the operator-defined quorum $v_{\min}$. Requiring a quorum rather than a single trigger makes the HPC decision robust to noisy or partially available metrics: a fragment is only forced to classical execution when multiple structural dimensions independently signal that it exceeds QPU capacity. If neither the QC envelope nor the HPC quorum is satisfied, the label is \texttt{Undecided}, and the router resolves it at dispatch time. All thresholds and $v_{\min}$ are set by the operator to reflect the QPU's qubit capacity, connectivity constraints, and depth tolerance.

\item \emph{Score.}
Each metric is normalized against the operator-defined QC upper bound and combined into a weighted \emph{QC pressure score}:
\begin{equation}\label{eq:score}
  S_i \;=\;
    w_q\,\frac{q_i}{q^{\max}_{\mathrm{QC}}}
  + w_d\,\frac{d_i}{d^{\max}_{\mathrm{QC}}}
  + w_g\,\frac{g_i}{g^{\max}_{\mathrm{QC}}}
  + w_z\,\frac{z_i}{z^{\max}_{\mathrm{QC}}},
\end{equation}
where the weights $w_q, w_d, w_g, w_z \ge 0$ sum to~1 and reflect the operator's assessment of which structural dimension most constrains the target QPU. Lower scores indicate stronger QC affinity. The label is then assigned by comparing $S_i$ against two operator-defined thresholds $\tau_{\mathrm{QC}}$ and $\tau_{\mathrm{HPC}}$ separated by a dead-band gap $\delta \ge 0$:
\begin{equation}\label{eq:score-decision}
  \ell_i =
  \begin{cases}
    \texttt{QC}        & \text{if } S_i \le \tau_{\mathrm{QC}} - \delta, \\
    \texttt{HPC}       & \text{if } S_i \ge \tau_{\mathrm{HPC}} + \delta, \\
    \texttt{Undecided} & \text{otherwise.}
  \end{cases}
\end{equation}
The dead band $[\tau_{\mathrm{QC}}{-}\delta,\,\tau_{\mathrm{HPC}}{+}\delta]$ absorbs fragments whose metrics are ambiguous, deferring their placement to the DQR capacity-aware resolver at dispatch time.

\item \emph{Hybrid.}
Budget rules~\eqref{eq:budget-qc} are applied first; if they yield a decisive \texttt{QC} or \texttt{HPC} label, that label is used. Fragments that remain \texttt{Undecided} under the budget pass are then re-evaluated by the score function~\eqref{eq:score}--\eqref{eq:score-decision}, which may resolve them or leave them \texttt{Undecided} if $S_i$ falls within the dead band. This two-pass structure lets operators encode hard hardware constraints in the budget layer while using a continuous score to resolve borderline cases.

\item \emph{Autobudget.}
When all fragments in a batch share nearly identical structural metrics, threshold policies tend to assign every fragment to the same class. Autobudget addresses this by enforcing a globally specified fraction triplet $(\alpha_{\mathrm{QC}},\, \alpha_{\mathrm{HPC}},\, \alpha_{\mathrm{U}})$, \review{normalized} to sum to~1, which the operator sets to reflect the desired HPC--QC workload split. Given $n$ fragments, the target counts are:
\begin{equation}\label{eq:autobudget-targets}
\begin{aligned}
  T_{\mathrm{QC}}  &= \left\lfloor \alpha_{\mathrm{QC}}\,n + \tfrac{1}{2} \right\rfloor,\\
  T_{\mathrm{HPC}} &= \left\lfloor \alpha_{\mathrm{HPC}}\,n + \tfrac{1}{2} \right\rfloor,\\
  T_{\mathrm{U}}   &= n - T_{\mathrm{QC}} - T_{\mathrm{HPC}}.
\end{aligned}
\end{equation}
The assignment proceeds in two passes. First, every fragment receives a baseline label from the budget rule~\eqref{eq:budget-qc} and a pressure score $S_i$ from~\eqref{eq:score}. Second, the pressure scores are used to rank fragments by QC affinity: those within the QC envelope are sorted by ascending $S_i$ and the top $T_{\mathrm{QC}}$ are assigned \texttt{QC}; the remaining fragments are sorted by descending $S_i$ and the top $T_{\mathrm{HPC}}$ are assigned \texttt{HPC}; all others become \texttt{Undecided}. This global reassignment guarantees quota compliance regardless of the threshold configuration, making it the appropriate choice when the fragment set is homogeneous ---or the hardware-based physical suitability is unknown--- and the operator wants explicit control over the QC-to-HPC dispatch ratio.
\end{itemize}

These labels serve as routing hints to the orchestration layer but are not binding: the router may override them based on capacity constraints, policies, and failure history. The labeling stage therefore decouples domain knowledge about hardware capabilities from the scheduling mechanism, allowing either to evolve independently.

\subsection{Runtime Orchestration Layer}\label{subsec:runtime-orchestration-layer}

The Dynamic Queue Router (DQR) is the core scheduling component developed for the orchestration layer. It maintains an in-memory context that tracks the state of every fragment across its execution lifecycle (see Figure~\ref{fig:state-machine} described in Section~\ref{sec:meth}). The DQR context is owned exclusively by the execution coordinator process (MPI rank~\#0). Worker MPI ranks \#(1..n-1) execute fragment simulations and communicate results back to rank~\#0 via MPI tags (\texttt{TAG\_DONE}, \texttt{TAG\_FAILED}). 
The coordinator-worker protocol is built on a minimal two-message exchange: rank~0 issues a \textsc{Work} message carrying the fragment identifier and routing metadata, and the receiving worker replies with \textsc{Done} or \textsc{Failed} upon completion. Result data is never transmitted over MPI; workers write expectation values directly to the GPFS, keeping message size bounded and independent of fragment complexity. Worker selection follows a ring traversal over ranks $1$ through $n-1$: the sender advances a circular pointer across the worker set, skipping occupied ranks, and assigns each fragment to the next free slot. This round-robin baseline avoids centralized queue contention and distributes load evenly under uniform fragment durations while remaining $\mathcal{O}(n)$ per wave ---negligible relative to execution latency.

\review{
MPI is used here only as a thin transport: all scheduling state resides in
the coordinator, no collective operations are used, and result data bypasses
MPI through GPFS. The coordinator relies on only three primitives: launching
$n$ workers under the resource manager, sending a work item to a specific
worker, and testing for completion without blocking. DQR is therefore not
tied to MPI, and any runtime providing these primitives, including
task-based runtimes such as StarPU~\cite{starpu}, PaRSEC~\cite{parsec},
HPX~\cite{hpx} or Charm++~\cite{charm}, could host it. We retain MPI because
Slurm launches MPI jobs natively, the capacity model reads the allocation
directly from the scheduler, no additional runtime is required on the
production system, and our malleability roadmap~\cite{iserte_mpi_2025, sandas_production_2026}
operates on MPI jobs.
}
 
DQR operates in discrete \emph{dispatch waves}. In each wave, the router applies its selection policy to the current set of \texttt{PENDING} fragments, produces a \emph{dispatch plan} listing the indices of fragments selected for HPC and QC backends, commits this plan to the fragment state machine (transitioning selected fragments from \texttt{PENDING} to \texttt{DISPATCHED}), and submits the corresponding simulation tasks to the backend sender. The wave counter is incremented on every commit, even for empty plans, so that iteration-dependent routing rules are evaluated consistently. 
At each tick, the coordinator reads the live backlog (\texttt{hpc\_pending}, \texttt{qc\_pending}, \texttt{undecided\_pending}, \texttt{hpc\_dispatched}, \texttt{qc\_dispatched}) alongside the current capacity snapshot and produces a routing plan that differs from the previous wave: HPC completions free slots that Undecided fragments can claim, QC tails hold slots that would otherwise go to HPC, and a permanent QC failure triggers an immediate re-label ---based on policy configuration--- and re-enqueue as HPC without stalling the pipeline.

In this layer, we can also find the resource manager system responsible for orchestrating HPC resources. 

In the current implementation, HPC resources are managed by Slurm~\cite{10.1007/10968987_3}, while the QC capacity is read from environment variables (\texttt{DQR\_QC\_SLOTS\_TOTAL}).

The wave-based dispatch mechanism is the mechanism through which the DQR achieves both parallelism and adaptivity following Algorithm~\ref{alg:wave} described in the previous section.

Rank~\#0 non-blocking polls completions with \texttt{MPI\_Iprobe} to collect those that have arrived since the previous wave without busily waiting.
This means that a new dispatch wave is planned and committed even while earlier fragments are still executing, achieving pipeline-level concurrency: the system maintains multiple in-flight fragments at all times, limited only by the available computational units (capacity).
 
The capacity is refreshed at each wave from the Slurm allocation environment variables (\texttt{SLURM\_NTASKS}, \texttt{SLURM\_NNODES}) and from DQR-specific environment variables (\texttt{DQR\_QC\_DEGRADED} and \texttt{DQR\_QC\_SLOTS\_TOTAL}). The number of available HPC slots is computed as the total number of MPI ranks minus one (the coordinator). 
When the QPU backend is degraded or offline, the \texttt{qc\_degraded} flag suppresses all QC routing until the flag is cleared.
 
The routing policy within a wave is self-aware and label-sensitive. 
In this regard, at wave~0, the router processes labeled fragments (\texttt{HPC} and \texttt{QC}) first and defers \texttt{Undecided} fragments to a second pass, where they are filled into remaining HPC or QC slots according to the \texttt{DQR\_PREFER\_ITER0\_UNDECIDED} preference. 
From wave~1 onward, the router prioritizes QC-labeled fragments first to preserve QPU scheduling invariants, then routes HPC--labeled and \texttt{Undecided} fragments in an order controlled by \texttt{DQR\_PREFER\_ITERN\_UNDECIDED}. This asymmetry reflects the fact that QPU sessions are the scarce, long-latency resource: QC fragments should be submitted early to minimize total makespan.
 
The failure handling is integrated directly into the wave loop. A transient failure ---e.g., a QCore timeout or a simulator error--- requeues the fragment as \texttt{PENDING} with its failure counter incremented; once the counter exceeds \texttt{DQR\_MAX\_TRANSIENT\_RETRIES}, the failure is promoted to permanent. A permanent failure on a QC-routed fragment can trigger a failover to HPC if the \texttt{DQR\_ALLOW\_FAILOVER\_QC\_TO\_HPC} flag is set in the policy, in which case the fragment is re-labeled at the routing level and re-enqueued as an HPC task. This mechanism ensures that QPU unavailability does not stall the reconstruction pipeline.

The fundamental property that makes this topology viable without checkpoints is the independence of fragments. Because no worker communicates with any other worker and no fragment depends on the output of a peer, rank~0 is never required to drain a wave before planning the next one. Completions from wave $k$ are harvested opportunistically during the planning phase of wave $k+1$ or later, meaning that at any given moment, in-flight fragments may belong to different logical waves. The wave counter therefore serves as a routing-policy index, not a global barrier: utilization remains bounded by aggregate worker capacity rather than by the latency of the slowest fragment in any cohort.

\subsection{Execution Backends}\label{subsec:backends}

The backend layer is encapsulated in QCore, a C++ library with a C API (\texttt{qcore\_run\_meta\_file\_ex}), developed for this purpose, which abstracts the execution targets. The target is selected by the \texttt{backend\_target} field injected into the fragment metadata by the DQR coordinator before dispatch.

The \emph{HPC backend} uses Qulacs for state vector simulation via a direct circuit construction from the 
OpenQASM 2.0. It supports the gate set $\{H,\, X,\, Z,\, \mathrm{CX},\, 
\mathrm{CZ},\,$ $R_x(\theta),\, R_y(\theta),\, 
R_z(\theta),\, U(\theta,\,\phi,\,\lambda)\}$, and computes expectation values by sampling the final state vector with a configurable shot count (\texttt{QCORE\_SHOTS}, by default 1024).
 
The \emph{QPU backend} invokes a Python bridge via \texttt{popen}. The bridge is implemented in \verb|backend_qpu_qulacs.py|, and submits the OpenQASM circuit to the CESGA Qmio QPU through the \texttt{qmio} SDK, polls for completion, and returns counts in the same JSON schema as the CPU backend. When the environment variable \texttt{Qmio\_DRY\_RUN=1} is set, the bridge substitutes a local Qulacs simulation, enabling end-to-end testing of the QPU execution path without access to physical hardware. 
 
The backends return responses conforming to the JSON schema \texttt{qcore.result.v1} , which includes the fragment expected value, shot count, Pauli string, backend identity, and per-phase timing information. The DQR coordinator parses the \texttt{expected\_value} field from each completion and passes it to the tensor reconstruction stage; no other field is required for correctness, ensuring that future adapters can be added without changes to the coordination or reconstruction logic.

\section{Performance Evaluation}\label{sec:eval}
We evaluate the DQR framework along two primary axes: execution time and resource utilization. Correctness of  the reconstruction pipeline is assessed separately in Section~\ref{subsec:validation}. Starting from a monolithic CPU baseline, we compare hybrid configurations that vary the labeling policy, the QC slot count, and the MPI process count, isolating the contribution of each degree of freedom on completion time and efficiency. The evaluation covers two complementary QPU regimes: local on-premises execution on the CESGA Qmio processor and remote execution via the IBM~Torino cloud service.

Along the experiments, we are going to use the following nomenclature to refer to timings:

\begin{itemize}
    \item QPU-time: time billed by the QPU backend for the actual execution of the logic gate, excluding queue waiting time in the remote QPU scenario. 
    \item DQR-time: execution time of the fragments in the HPC and QC resources, covering fragment dispatch across HPC and QPU-time plus the tensor reconstruction. 
    \item Makespan: elapsed time between job initialization and completion, excluding Qmio queue waiting time. 
    Depending on the type of job, the makespan represents the times for:
    \begin{itemize}
        \item QPU-only:\\
                $t_{\text{setup}_q} + t_{\text{round-trip}} + t_{\text{QPU-time}} + t_{\text{post-process}}$
        \item Hybrid:\\
                $t_{\text{setup}_h} + \max(t_{\text{round-trip}},\ t_{\text{DQR-time}})$
    \end{itemize}
    where $t_{\text{setup}_q}$ covers circuit loading, circuit mapping, transpilation, and expectation value for QPU-only jobs; while 
    $t_{\text{setup}_h}$ environment setup, gRPC server startup, circuit cutting, circuit labeling, MPI host resolution, and artifact submission for hybrid jobs.
    When using an off-premises QPU, $t_\text{round-trip}$ corresponds to the QPU provider overhead---QPU queue wait, data network exchange, and job scheduling. For on-premises, this overhead  corresponds only to the QPU queue waiting time. In hybrid executions, since QPU and HPC fragments run concurrently, the maximum between $t_\text{round-trip}$ (QPU-related) and DQR-time (HPC--related) determines the fragment's total execution time.
    Finally, $t_\text{post-process}$ covers result retrieval and expected value reconstruction---the hybrid reconstruction remains within DQR-time.
\end{itemize}

Table~\ref{tab:metrics} summarizes these and the rest of the metrics leveraged along the performance evaluation of this paper.

\begin{table*}
\centering
\caption{Evaluation metrics and derived quantities.}
\label{tab:metrics}
\setlength{\tabcolsep}{4pt}
\renewcommand{\arraystretch}{1.02}
\small
\begin{tabularx}{\textwidth}{@{}L{0.1\textwidth} L{0.28\textwidth} Z@{}}
\toprule
\textbf{Symbol} & \textbf{Definition} & \textbf{Description} \\
\midrule

\rowcolor{fgcs@header}
\multicolumn{3}{@{}l}{\textit{Per-fragment timings}} \\

$t_{\text{setup}_q}$ &
$t_{\text{load}} + t_{\text{map}} + t_{\text{transpile}} + t_{\text{ev}}$ &
Pipeline overhead for QPU-only jobs: circuit loading, qubit mapping, transpilation, and observable. \\

$t_{\text{setup}_h}$ &
$t_{\text{env\_setup}} + t_{\text{gRPC}} + t_{\text{cut}} + t_{\text{sub}} + t_{\text{label}} + t_{\text{MPI\_host\_res}}$ &
Pipeline overhead for hybrid jobs: environment setup, gRPC server startup, circuit cutting, circuit labeling, MPI host resolution, and artifact submission for hybrid jobs. \\

$n_{\text{QC}}$ &
--- &
Number of fragments labeled QC in a given run. \\

$n_{\text{HPC}}$ &
--- &
Number of fragments labeled HPC in a given run. \\

$t_{\text{round-trip}}$ &
$t_{\text{communication}} + t_{\text{QPU\_queue\_waiting}} + t_{\text{job scheduling}}$ &
Corresponds to the QPU provider overhead---QPU queue wait, data network exchange, and job scheduling. For on-premises, this overhead  corresponds only to the QPU queue waiting time. \\

$\bar{t}_{\text{QPU}}$ &
$\dfrac{1}{n_{\text{QC}}}\displaystyle\sum_{i=1}^{n_{\text{QC}}} t_{\text{round-trip}}$ &
Mean round-trip time over all $n_{\text{QC}}$ QC-labeled fragments in a run. \\

$\bar{t}_{\text{HPC}}$ &
$\dfrac{1}{n_{\text{HPC}}}\displaystyle\sum_{i=1}^{n_{\text{HPC}}} t_{\text{exec}}$ &
Mean execution time over all $n_{\text{HPC}}$ HPC--labeled fragments in a run. \\

\midrule
\rowcolor{fgcs@header}
\multicolumn{3}{@{}l}{\textit{Critical-path times}} \\

$T_{\text{QC}}$ &
$\max_{i \in \text{QC}}\bigl(t_{\text{done}}\bigr) - t_{\text{start}}$ &
Elapsed time from first dispatch to last QC fragment completion; equals $\lceil n_{\text{QC}}/s \rceil \times \bar{t}_{\text{QPU}}$ under uniform round-trips, where $s$ is the QC slot count. \\

$T_{\text{HPC}}$ &
$\max_{i \in \text{HPC}}\bigl(t_{\text{done}}\bigr) - t_{\text{start}}$ &
Elapsed time from first dispatch to last HPC fragment completion. \\

\midrule
\rowcolor{fgcs@header}
\multicolumn{3}{@{}l}{\textit{Aggregate times}} \\

$\text{Makespan}_{\text{CPU}}$ &
$t_{\text{CPU\_sim}} + t_{\text{sampling}}$ & Elapsed time between job initialization and completion for the CPU-only baseline. Comprises the Qulacs statevector simulation and shot sampling with expectation-value computation.\\

$\text{Makespan}_{\text{DQR}}$ &
$t_{\text{setup}_h} + \max\!\bigl(t_{\text{round-trip}},\,
    \text{DQR-time}\bigr)$ & Elapsed time between job initialization and completion for a hybrid DQR job, excluding Qmio queue waiting time. For QPU-only jobs, the analogous decomposition is $t_{\text{setup}_q} + t_{\text{round-trip}} + t_{\text{QPU-time}} + t_{\text{post-process}}$.\\

DQR-time &
$\max(T_{\text{QC}},\,T_{\text{HPC}}) + C_{\text{fixed}}$ &
Elapsed time from MPI initialization to last reconstruction within DQR; includes fragment dispatch, QPU round-trips, and tensor reconstruction. \\

$C_{\text{fixed}}$ &
$t_{\text{DQR-time}} - \max(T_{\text{QC}},\,T_{\text{HPC}})$ &
Fixed coordination residual: MPI communication, rank-zero polling, and collective overhead not on the QC or HPC critical path. \\

Speedup &
$\text{Makespan}_{\text{CPU}} / \text{Makespan}_{\text{DQR}}$ &
Ratio of CPU-baseline makespan to DQR makespan; values ${>}1$ indicate improvement. \\


\midrule
\rowcolor{fgcs@header}
\multicolumn{3}{@{}l}{\textit{Reconstruction observable}} \\

$\langle O \rangle$ &
$\displaystyle\sum_i c_i \langle O_i \rangle$ &
Reconstructed expectation value of the observable; $c_i$ are the quasi-probability coefficients from the gate-cut decomposition~\cite{mitarai_constructing_2021}, and $\langle O_i \rangle$ is the expectation value estimated from fragment~$i$. \\

\midrule
\rowcolor{fgcs@header}
\multicolumn{3}{@{}l}{\textit{Balance and overlap metrics}} \\

$\Phi$ &
$T_{\text{QC}} / T_{\text{HPC}}$ &
DQR balance number: $\Phi > 1$ is QC-bound, $\Phi < 1$ is HPC--bound, $\Phi = 1$ is balanced. \\

$\sigma$ &
$\begin{aligned}
\sigma &= \frac{T_{\text{QC}} + T_{\text{HPC}}}{\max(T_{\text{QC}},\,T_{\text{HPC}})} \\
       &= 1 + \frac{\min(T_{\text{QC}},\,T_{\text{HPC}})}{\max(T_{\text{QC}},\,T_{\text{HPC}})}
\end{aligned}$ &
Overlap speedup: ratio of sequential to concurrent dispatch time; $\sigma \in (1,\,2]$, with $\sigma = 2$ iff $\Phi = 1$. \\

\bottomrule
\end{tabularx}
\end{table*}

\subsection{Experimental setup configuration}\label{subsec:setup}
The experiments are executed on the CESGA Qmio cluster~\footnote{\url{https://cesga-docs.gitlab.io/qmio-user-guide}}, where each compute node is equipped with 64~cores and 1~TB of RAM. 
This cluster integrates an on-premises Oxford Quantum Circuit’s superconducting QPU. 
The QPU consists of 32 coaxmon qubits, and it is handled by microwave pulses crafted by QAT Software~\footnote{\url{https://github.com/oqc-community/qat}}.
Furthermore, Qmio provides the Qmio-Qulacs quantum emulation on CESGA infrastructure, supported by hardware and software based on a distributed version of Qulacs~\cite{Suzuki2020QulacsAF} running on Fujitsu FX700 machines. \review{DQR does not use this service: its own Qulacs backend (QCore, Table~\ref{tab:software_stack}) runs directly on the general-purpose Qmio compute nodes described above (64 cores, 1~TB RAM each), independently of the Fujitsu~FX700 partition.}
Additionally, we have connected the cluster to the remote QPU IBM~Torino, via the IBM Quantum Cloud service~\footnote{\url{https://quantum.cloud.ibm.com}}. IBM~Torino is powered by the Heron~r1 processor~\cite{10.1007/s11227-025-07047-7}. A 133-qubit superconducting chip arranged in a heavy-hexagonal lattice.
QCore Backends for both QPUs are implemented and integrated into DQR as described in Section~\ref{subsec:backends}.

Regarding the software stack, Table~\ref{tab:software_stack} summarizes what we have utilized to perform the experiments.

\begin{table}
\centering
\caption{Software summary. The \textit{Stack} column indicates if the component is exclusive for the DQR framework or it is also used in the QPU-only scenario (\textit{Both}).}
\label{tab:software_stack}
\setlength{\tabcolsep}{3pt}
\renewcommand{\arraystretch}{1.05}
\resizebox{\columnwidth}{!}{%
\begin{tabular}{@{}llll@{}}
\toprule
\textbf{Component} & \textbf{Version} & \textbf{Role} & \textbf{Stack} \\
\midrule
\rowcolor{fgcs@header}
\rowcolor{fgcs@header}
\multicolumn{4}{@{}l}{\textit{Compiler \& Build Tools}} \\
GCC              & 12.3.0           & C/C++ compiler (QCore, RMS)         & Both \\
LLVM             & 16.0.0           & Compiler infrastructure (Qulacs)    & Both \\
Python           & 3.9.9            & Interpreter (QPU bridge, QCut)      & Both \\
\midrule
\rowcolor{fgcs@header}
\multicolumn{4}{@{}l}{\textit{HPC}} \\
SLURM            & 23.11.4          & Resource manager    & Both \\
Open MPI/PRRTE         & 6.1.0a1          & MPI runtime  & DQR  \\
UCX              & 1.11.2           & MPI transport layer (TCP) & DQR \\
\midrule
\rowcolor{fgcs@header}
\multicolumn{4}{@{}l}{\textit{QC}} \\
qmio-run         & 0.5.1            & CESGA Qmio runtime module           & Both \\
qmio-tools       & 0.2.1            & \texttt{QmioBackend} / QPU SDK      & Both \\
Qiskit           & 2.2.3            & Quantum circuit compilation         & Both \\
qiskit-qasm3-import & 0.6.0         & QASM~3.0 parser for Qiskit          & Both \\
qiskit-ibm-runtime & 0.46.1         & IBM Quantum Cloud job submission   & Both \\
Qdislib          & 1.0.0            & Circuit cutting (\texttt{find\_cut}) & DQR \\
Qulacs           & 0.6.13              & CPU state vector simulator (QCore)   & DQR \\
\bottomrule
\end{tabular}%
}
\end{table}

We evaluate the proposed framework on a 32-qubit circuit.
Particularly, a Hardware-Efficient Ansatz (HEA) with one layer ($L{=}1$) is chosen as the  benchmark circuit for three main reasons: its regular $\mathrm{CZ}$ entanglement pattern produces predictable cut locations, making it a standard HPC--QC benchmark~\cite{10.1145/3731599.3767547}; its $Z^{\otimes n}$ expectation value concentrates near zero for randomly-parametrized instances, providing an analytically grounded correctness check; and the accuracy of the reconstructed observable is \emph{not} the primary concern of this evaluation---it is determined by the cutting framework and reconstruction method, both external to DQR. The HEA circuit is therefore used exclusively as a controlled vehicle to exercise the orchestration layer---scheduling behavior, resource utilization, failure handling, and dispatch latency---under realistic HPC--QC conditions.

\review{Because the DQR coordinator is label-agnostic and never inspects circuit code (Section~\ref{subsec:imp_cut}), the dispatch-level behavior evaluated here (slot contention, multi-wave replanning, failover and coordination overhead) does not depend on the algorithm that produced the parent circuit. Other algorithms change the distribution of fragment descriptors rather than the dispatch code path. We do not include random circuits: their unstructured entanglement mainly increases the number of cuts and the sampling overhead of the cutting stage, which is external to DQR.}

In this regard, the HEA circuit comprises 96~gates---32 two-qubit $\mathrm{CZ}$ and 64 single-qubit $R_Y$/$R_Z$---at depth 34. Parameters $\boldsymbol{\theta}$ are drawn uniformly from $[-\pi,\pi]$ 
via \texttt{qibochem.ansatz.he\_circuit}~\footnote{\url{https://qibo.science/qibochem/stable/api-reference/ansatz.html}}. With $k{=}2$ $\mathrm{CZ}$ gate cuts, Qdislib produces $6^2{=}36$ reconstruction terms across two circuit components, for a total of 72 subcircuit evaluations per run. To probe behavior at greater circuit depth, Table~\ref{tab:l2_comparison} repeats the evaluation at $L{=}2$ ($192$ gates, depth $68$, $k{=}4$, $2,592$ evaluations).


In order to make efficient use of the system resources, the experiments are designed to balance the load fairly among the requested computational units. 

The experiments are designed to operate in a \emph{mixed-latency regime}: HPC fragments complete in under a second, while QC fragments take several seconds each, so the MPI coordinator must continuously reschedule---dispatching Undecided fragments, absorbing failures, and rerouting through failover---until the last fragment completes. This is precisely the regime where the wave-based dispatch demonstrates its value: multi-wave replanning, QC slot contention, and the failover path are all exercised under realistic contention. A configuration with one MPI rank per fragment would collapse the workload into a single wave, eliminating these dynamics entirely. 

HPC and QC resources are provisioned to maintain load balance and minimize resource idle time. The ratio of HPC ranks to QC slots is tuned so that HPC and QC paths remain simultaneously active across multiple waves: neither backend exhausts its queue before the other, keeping the coordinator continuously replanning. A configuration with, say, 72 workers and 72 QC slots would collapse the entire workload into a single wave, eliminating multi-wave replanning, Undecided arbitration, and failover events from the observable trace entirely. The chosen allocation is therefore the minimal one that keeps all three routing paths---HPC, QC, and Undecided-to-either---simultaneously active across multiple waves, making the load-balancing and adaptive-dispatch behavior of the DQR fully observable and measurable throughout the experiment. Specific allocations per configuration are detailed in Table~\ref{tab:summary}.

\subsection{On/Off-premise QPUs}
Following, the presented infrastructure is evaluated twofold, with the Qmio local QPU and the remote QPU hosted in the cloud.

\subsubsection{Local QPU}\label{subsubsec:local_qpu}
Under fully on-premises conditions, firstly we run the baseline, which is the full uncut 32-qubit HEA circuit executed directly on the QPU.
Then, we evaluate the DQR-enabled scenario to study how the circuit fragments are routed between the QPU and the HPC resources.

In the baseline, the Qiskit transpilation expands the circuit from depth 35 to depth 374 and from 128 to 795 gates to conform to the native gate set and connectivity. 
The Qmio QPU completes the execution in $9.14$ seconds, for a total makespan of $9.75$ seconds, which adds the transpilation time.
This result serves as the local hardware reference: the local QPU executes the full 32-qubit circuit in under 10 seconds, a timing that the \textit{state vector} simulation on a single HPC node cannot match at this qubit count.
 
The hybrid execution applies the DQR pipeline with \texttt{DQR\_QC\_BACKEND=local}, routing QC-labeled fragments to the Qmio QPU. 
Of the 72 subcircuits produced by Qdislib ($k{=}2$ gate cuts), the \textit{Autobudget} labeler assigns seven fragments to the QC backend, 58 to the HPC, and the remaining 7 Undecided. 
The DQR dispatcher serializes local QPU dispatch, submitting one fragment at a time, as the Qmio backend does not expose a parallel slot count through the environment interface used by the capacity model. 
In the experiment, five out of the seven QC-labeled fragments, completed successfully on the Qmio QPU, confirming that the hardware can execute gate-cut subcircuits. 
The remaining two fragments 
fail at runtime with error \texttt{QCORE\_ERUNTIME}: \textit{``The control-flow construct \texttt{if\_else} is not supported by the backend.''} 
This failure arises because Qdislib's gate-cut decomposition injects mid-circuit classical~\texttt{if\_else} branches that the Qmio native stack cannot compile. 
Both fragments exhaust the single allowed retry, transition to \texttt{PERMANENT\_FAIL}, and are immediately relabeled and reenqueued as HPC tasks via the \texttt{DQR\_ALLOW\_FAILOVER\_QC\_TO\_HPC} policy. 
The failover is fully transparent: the pipeline completes all 72 fragments and performs tensor reconstruction without user intervention, yielding a makespan of $44.6$~s, from which $27.2$~s of DQR-time.

Table~\ref{tab:local_eval} compiles the configuration of the experiments and their results.
The \textit{Configuration} defines for each environment, respectively, the circuit features; the percentages of fragments labeled QC, HPC, and Undecided for DQR; MPI ranks for the HPC backend of DQR, which always needs an additional rank for the orchestration; and the number of HPC nodes (at least one for communicating with the QPU) to spread uniformly the MPI ranks.

The \textit{Quantum circuits} area accounts for the number of circuits (or fragments in DQR) completed. There are two fragments that, after failing in the QC environments, were re-routed to the HPC to be executed.

\textit{Performance} presents the timings for the different metrics as defined in Section~\ref{sec:eval}. More details are provided for the execution fragment to present the variability they present in terms of \textit{Mean/Median/Range}.

Finally, in \textit{Results}, reports the reconstructed expectation value $\langle O\rangle$; its interpretation against the CPU baseline and the exact state vector reference is discussed in
Section~\ref{subsec:validation}.

\begin{table}
\centering
\caption{Local QPU evaluation on the CESGA Qmio cluster.}
\label{tab:local_eval}
\setlength{\tabcolsep}{4pt}
\renewcommand{\arraystretch}{1.10}
\resizebox{\columnwidth}{!}{
\begin{tabular}{@{}lrr@{}}
\toprule
\textbf{Metric}
  & \textbf{QPU-only}
  & \textbf{Hybrid} \\
\midrule
 
\rowcolor{fgcs@header}
\multicolumn{3}{@{}l}{\textit{Configuration}} \\
Circuit ($n$, gates, depth)      & 32\,q, 128, 35     & 32\,q, 96, 34 \\
QC|HPC|U (\%)                     & ---                & 10\%\,|\,80\%\,|\,10\% \\
QC slots                              & ---                  & 1 \\
MPI ranks (\textit{np})           & ---                & 12+1 \\
HPC Nodes                             & 1                  & 2 \\
 
 
\midrule
\rowcolor{fgcs@header}
\multicolumn{3}{@{}l}{\textit{Quantum circuits}} \\
\# Completed            & 1             & 5 of 7 fragments\\
\# Failed: QC $\to$ HPC & ---           & 2 of 7 fragments\\

\midrule
\rowcolor{fgcs@header}
\multicolumn{3}{@{}l}{\textit{Performance}} \\
QPU-time      & $9.136$ s    &  ---\\
Fragments (65+2)     & --- & $6.6 \pm 4.7$~s/4.5~s/3.8--16~s  \\
DQR-time                    & ---                & $27.2$ s \\
Makespan     & $9.8$ s       & $44.6$ s \\
 
\midrule
\rowcolor{fgcs@header}
\multicolumn{3}{@{}l}{\textit{Results}} \\
$\langle O\rangle$                & $+0.023$           & $-6.6{\times}10^{-3}$ \\
 
\bottomrule
\end{tabular}
}
\end{table}

In conclusion, the hybrid run is slower than the QPU-only baseline for two reasons that are independent from the DQR architecture. 
First, the pipeline processes 72 fragments with 12 MPI ranks, so even the purely classical portion requires multiple dispatch rounds. 
Second, the two re-routed QC $\to$ HPC fragments each add approximately 7--9 seconds to the critical path, serialized because the capacity model receives no slot count from the local QPU backend interface. 
Despite these constraints, the hybrid run demonstrates that the failover mechanism functions correctly under a real QPU rejection condition: the DQR state machine recovers from hardware-level errors at the fragment granularity and preserves pipeline integrity without global restart.

Enabling local QPU execution of cut circuits would require either a control-flow-aware native compiler for Qmio or an alternative cutting strategy that avoids mid-circuit classical branches, a direction out of the scope of this work.

\subsubsection{Remote QPU}
The local evaluation has exposed a fundamental constraint that prevents using the Qmio QPU for the full cutting pipeline: the hardware does not support the classical control-flow inserted by Qdislib's gate-cut decomposition. This limitation is architectural rather than a deficiency of the DQR framework, and for this reason, an additional QPU could be used.
Particularly, in this section we target the IBM Quantum Cloud service, which compiles and executes the required control-flow constructs, natively. 

To start with, as in the local QPU scenario, the baseline is determined by executing the full uncut 32-qubit HEA circuit on the IBM~Torino QPU.
Then, in the hybrid approach, the same circuit decomposed into 72 fragments via the DQR pipeline is executed, routing seven QC-labeled fragments to IBM~Torino across three cloud slots and the remaining 65 fragments to CESGA HPC nodes due to system state and pending queue. Results are reported in Table~\ref{tab:remote_eval}.

The QPU-only run transpiles the 32-qubit logical circuit to 133 physical qubits on the IBM~Torino 133-qubit heavy-hex topology, expanding from depth 35 to depth 37 and from 128 to 192 native gates. 
Table~\ref{tab:remote_eval} summarizes the configuration and showcases the experimental results.
The same seven QC-labeled fragments are successfully completed with IBM~Torino.
In this scenario, $\bar{t}_{\text{QPU}}$ is substantially larger than in the local scenario: for off-premises execution $t_{\text{round-trip}}$ includes not only QPU-time but also QPU queue wait, data network exchange, and job scheduling---overhead that is virtually null for on-premises QPUs. The higher $\bar{t}_{\text{QPU}}$ increases $T_{\text{QC}}$ and therefore DQR-time and makespan relative to the local QPU setup.

 
The $\langle O\rangle$ values ($+3.4\times10^{-3}$ QPU-only, $-2.7\times10^{-5}$ hybrid) are both consistent with zero within the sampling noise floor at 1024 shots ($\sigma \approx 0.031$ for $\langle Z^{\otimes 32}\rangle$ on a randomly-parametrized HEA), so neither result carries statistical significance, and no systematic bias attributable to the cutting or reconstruction procedure can be identified from this comparison.

\begin{table}
\centering
\caption{Remote QPU evaluation on the CESGA Qmio cluster.}
\label{tab:remote_eval}
\setlength{\tabcolsep}{4pt}
\renewcommand{\arraystretch}{1.10}
\resizebox{\columnwidth}{!}{
\begin{tabular}{@{}lrr@{}}
\toprule
\textbf{Metric}
  & \textbf{QPU-only}
  & \textbf{Hybrid} \\
\midrule
 
\rowcolor{fgcs@header}
\multicolumn{3}{@{}l}{\textit{Configuration}} \\
Circuit ($n$, gates, depth)          & 32\,q, 128, 35       & 32\,q, 96, 34 \\
QC|HPC|U (\%)                         & ---                  & 10\%\,|\,80\%\,|\,10\% \\
QC slots                              & ---                  & 3 \\
MPI ranks (\textit{np})               & 1                    & 12+1 \\
HPC Nodes                                 & 1                    & 2 \\
 
\midrule
\rowcolor{fgcs@header}
\multicolumn{3}{@{}l}{\textit{Quantum circuits}} \\
\# Completed                & 1    & 7 of 7 fragments \\
\# Failed: QC $\to$ HPC    & ---                 & --- \\
 

\midrule
\rowcolor{fgcs@header}
\multicolumn{3}{@{}l}{\textit{Performance}} \\
QPU-time                        & 19 s                   & --- \\
Fragments                       & ---                  & $12.9 \pm 1.4$ s/\,13.1 s/\,11.4--14.7 s \\
DQR-time                        & ---                  & $41.7$ s \\
Makespan                        & $34.1$ s       & $50.7$ s \\
 
\midrule
\rowcolor{fgcs@header}
\multicolumn{3}{@{}l}{\textit{Results}} \\
$\langle O\rangle$                    & $+3.4{\times}10^{-3}$ & $-2.7{\times}10^{-5}$ \\
 
\bottomrule
\end{tabular}
}
\end{table}

\subsection{Results}\label{subsec:validation}
We evaluate DQR on the same 32-qubit HEA circuit introduced in Section~\ref{subsec:setup}, comprising 96 gates and depth~34, under a CPU-only baseline and four DQR dispatch policies (A--D). The labeling policy is specified as a QC/HPC/Undecided percentage triplet, which determines how the 72 fragments are classified before entering the dispatch loop. QC-labeled fragments are submitted to the remote IBM~Torino QPU, while HPC-labeled fragments run on the Qulacs simulator on the Qmio cluster. Undecided fragments are dynamically assigned by DQR at dispatch time, based on resource availability, filling QC and HPC slots according to the backend preference configured in the policy---independently for wave~0 and subsequent waves (see Section~\ref{subsec:runtime-orchestration-layer}).

Table~\ref{tab:summary} summarizes the five studied configurations: a monolithic CPU reference and four DQR runs under policies~A--D. In all cases, the observable $\langle Z^{\otimes 32}\rangle$ is estimated via finite-shot sampling with 1024 shots. The appropriate reference is therefore the CPU shot estimate $-0.014$ with sampling uncertainty $\sigma_{\mathrm{CPU}}\approx 0.031$, rather than the exact \texttt{state vector} value. Comparing against the exact reference $\langle Z^{\otimes 32}\rangle_{\mathrm{exact}} = 1.82\times 10^{-10}$ would conflate sampling noise with systematic error. The exact value simply confirms that the observable is essentially zero for this instance, so all deviations across configurations are sampling fluctuations. The CPU baseline itself lies only $0.44\,\sigma$ from the exact value. All DQR runs produce estimates consistent with the CPU reference (Table~\ref{tab:summary}, \textit{Results} section): the four IBM-remote policies (A--D) range from $-7.2\times 10^{-4}$ to $+1.8\times 10^{-3}$ (all $<0.06\,\sigma$), and the local Qmio hybrid yields $-6.6\times 10^{-3}$ ($0.21\,\sigma$).

The reconstruction metadata is identical across all four DQR configurations in Table~\ref{tab:summary}: 72 evaluations, 36 terms, 2 components, and a global factor of 0.25, matching the $6^2 = 36$ gate-cut terms at $k = 2$. This agreement shows that cutting, backend execution, and reconstruction are carried out consistently and that the DQR-time layer introduces no detectable systematic bias. Observable accuracy is governed by the underlying cutting framework and reconstruction method~\cite{mitarai_constructing_2021}, which are external to DQR itself.

\begin{table}
\centering
\caption{Summary comparison of CPU-only simulation (Qulacs) versus the
         DQR pipeline under four dispatch policies.}
\label{tab:summary}

\setlength{\tabcolsep}{2pt}
\renewcommand{\arraystretch}{1.05}

\resizebox{\columnwidth}{!}{
\begin{tabular}{@{}lrrrrr@{}}
\toprule
\textbf{Metric}
  & \textbf{CPU}
  & \textbf{Pol.~A}
  & \textbf{Pol.~B}
  & \textbf{Pol.~C}
  & \textbf{Pol.~D} \\
\midrule

\rowcolor{fgcs@header}
\multicolumn{6}{@{}l}{\textit{Configuration}} \\
QC/HPC/U (\%)
  & ---     & 10/80/10 & 20/60/20 & 20/60/20 & 30/50/20 \\
QC slots
  & ---     & 3  & 3  & 20 & 72 \\
MPI ranks (\textit{np})
  & ---     & 13 & 13 & 13 & 73 \\
HPC Nodes
  & 1       & 2  & 2  & 2  & 4 \\

\midrule
\rowcolor{fgcs@header}
\multicolumn{6}{@{}l}{\textit{Performance}} \\
CPU simulation (s) & 40.3 & --- & --- & --- & --- \\
Sampling + EV (s)  & 15.5 & --- & --- & --- & --- \\
Makespan (s)   & 56.2  & 50.7 & 91.0 & 59.6  & 54.9 \\ 
DQR-time (s) 
  & ---   & 41.7 & 76.7 & 51.7 &  45.6  \\
$T_{\text{QC}}$ (s) & --- & 37.2 & 72.4 & 47.0 & 39.8 \\
$T_{\text{HPC}}$ (s) & --- & 17.9 & 14.8 & 49.2 & 3.5 \\
$\bar{t}_{\text{QPU}}$ (s)
  & --- & $13.0{\pm}1.2$ & $15.0{\pm}3.1$ & $24.6{\pm}6.0$ & $26.2{\pm}8.4$ \\
$\bar{t}_{\text{HPC}}$ (s)
  & --- & $3.57{\pm}0.13$ & $3.48{\pm}0.14$ & $3.60{\pm}0.12$ & $3.22{\pm}0.11$ \\

\midrule
\rowcolor{fgcs@header}
\multicolumn{6}{@{}l}{\textit{Results}} \\
$\langle O\rangle$
  & $-0.014$
  & $-2.7{\times}10^{-5}$
  & $+1.8{\times}10^{-3}$
  & $-2.7{\times}10^{-4}$
  & $-7.2{\times}10^{-4}$ \\
\rowcolor{fgcs@best}
Speedup & --- & $1.11\times$ & \review{$1.62\times$ slower} & \review{$1.06\times$ slower} & $1.02\times$ \\
\bottomrule
\end{tabular}
}
\end{table}
\review{Policies A--D also provision more HPC nodes and MPI ranks than the CPU-only reference run (Table~\ref{tab:summary}); the reported speedups should therefore be read as time-to-solution improvements under an expanded resource budget, not as an efficiency-per-node comparison.}

\subsubsection{CPU baseline and fragmentization.}
The baseline executes the full 32-qubit HEA circuit as a single Qulacs \texttt{state vector} simulation using 64 OpenMP threads on a node. This requires allocating a $2^{32}$--entry complex \texttt{state vector} (64~GB of RAM). The simulation phase alone takes 40.3~s, and with sampling (1024 shots over the full 32-qubit $Z^{\otimes 32}$ observable), a further 15.5~s (Table~\ref{tab:summary}); the makespan of 56.2~s includes environment initialization overhead. In contrast, all DQR configurations decompose the circuit into 72 fragments of 16~qubits each, 
\review{
reducing the state-vector memory from 64~GB ($2^{32}\times16$~B) to
$2^{16}\times16$~B $=1$~MiB per fragment, i.e.\ under 2~MB per
MPI worker rank (excluding the fixed footprint of the Python, MPI and Qulacs
runtimes).}

\review{
The monolithic baseline is itself bounded by node memory: at 36 qubits the
state vector would require $2^{36} \times 16$~B $= 1$~TiB, the entire memory
of a Qmio node. Distributed state-vector simulation removes this bound at the
cost of inter-node communication, which depends on how qubits are mapped to
nodes~\cite{arejula_placement_2026}. The 32-qubit circuit therefore lies close to the largest size
at which a single-node monolithic reference is still possible, and the
speedups reported below should be read as a comparison at the edge of the
monolithic regime rather than in the regime DQR targets, where fragments
remain at about 1~MB per worker rank regardless of the size of the parent
circuit.
}

\paragraph{Policy~A, the optimal approach.}
Policy~A achieves the best absolute makespan among the DQR runs at 50.7~s, a $1.11\times$ improvement over the baseline. Its 10/80/10 labeling policy routes 58 of the 72 fragments to HPC and 7 to IBM~Torino, served by 3 concurrent QC slots across 12 worker ranks plus the coordinator. The 7 QC fragments therefore span $\lceil 7/3 \rceil = 3$ effective QPU rounds; at $\bar{t}_{\text{QPU}} = 13.0$~s per fragment (QPU-time $\approx 2$~s), $T_{\text{QC}} = 37.2$~s. Policy~A is QC-bound ($\Phi = T_{\text{QC}}/T_{\text{HPC}} = 2.08 > 1$): $T_{\text{HPC}} = 17.9$~s completes within the QC critical path, so HPC execution is fully overlapped with in-flight QC fragments and does not extend DQR-time $= 41.7$~s beyond the makespan.

\paragraph{Policy~B and slot-induced serialization.}
Policy~B delivers the worst performance, with a makespan of 91.0~s. Doubling the QC-labeled fraction relative to Policy~A sends 14~fragments to IBM~Torino, still over only 3~QC slots; these 14~fragments serialize into $\lceil 14/3 \rceil = 5$ QPU rounds of roughly 15~s each. Wave-level synchronization amplifies this cost: MPI ranks quickly exhaust their local shard batches, and subsequent waves carry an $T_{\text{QC}}$-induced backlog that leaves HPC capacity idle while QC slots drain. Increasing the fraction of QC labels without proportionally scaling the slot count thus degrades performance via a staircase serialization effect.

\paragraph{Policy~C and the effect of QC slots.}
Policy~C retains the same labeling policy as Policy~B (20/60/20, 13~MPI ranks) but increases the QC slot count from 3 to 20. With 20 concurrent slots, all 14 QC-labeled fragments fit within a single effective QPU round ($\lceil 14/20 \rceil = 1$), eliminating the staircase serialization that dominated Policy~B. In addition, the available slots allow DQR to resolve 15 Undecided fragments to QC at runtime (raising $n_{\text{QC}} = 29$; $\bar{t}_{\text{QPU}} = 24.6 \pm 6.0$~s per fragment; $\lceil 29/20 \rceil = 2$ effective rounds; $T_{\text{QC}} = 47.0$~s). Despite twice as many QC fragments, $T_{\text{QC}}$ drops from 72.4~s to 47.0~s because the slot count collapses five serialized rounds into two concurrent ones. The makespan drops from 91.0~s to 59.6~s, a 34.5\% reduction, confirming that the bottleneck in Policy~B was slot-induced QPU serialization rather than per-fragment $\bar{t}_{\text{QPU}}$. The Policy~B-versus-C comparison isolates QC slot count as the decisive variable for both makespan and HPC utilization under a fixed labeling policy.

Figure~\ref{fig:conceptual_frag_dispatch} illustrates this mechanism on a simplified scenario (10 fragments: 6 HPC, 4 QC; 2 MPI ranks). With 3 QC slots (analogous to Policy~B), the fourth QC fragment must wait for a slot to free, causing MPI workers to idle once their HPC queues empty. With 6 slots (analogous to Policy~C), all QC fragments are dispatched in a single wave, halving makespan and eliminating slot-induced queueing. In both cases, it is the slot count---not the rank count---that determines whether the QPU path is serialized into multiple rounds or clears in one.


\textbf{\begin{figure*}[htbp]
    \centering
    \includegraphics[clip,width=0.8\linewidth,trim={0cm 0cm 0cm 1.5cm}]{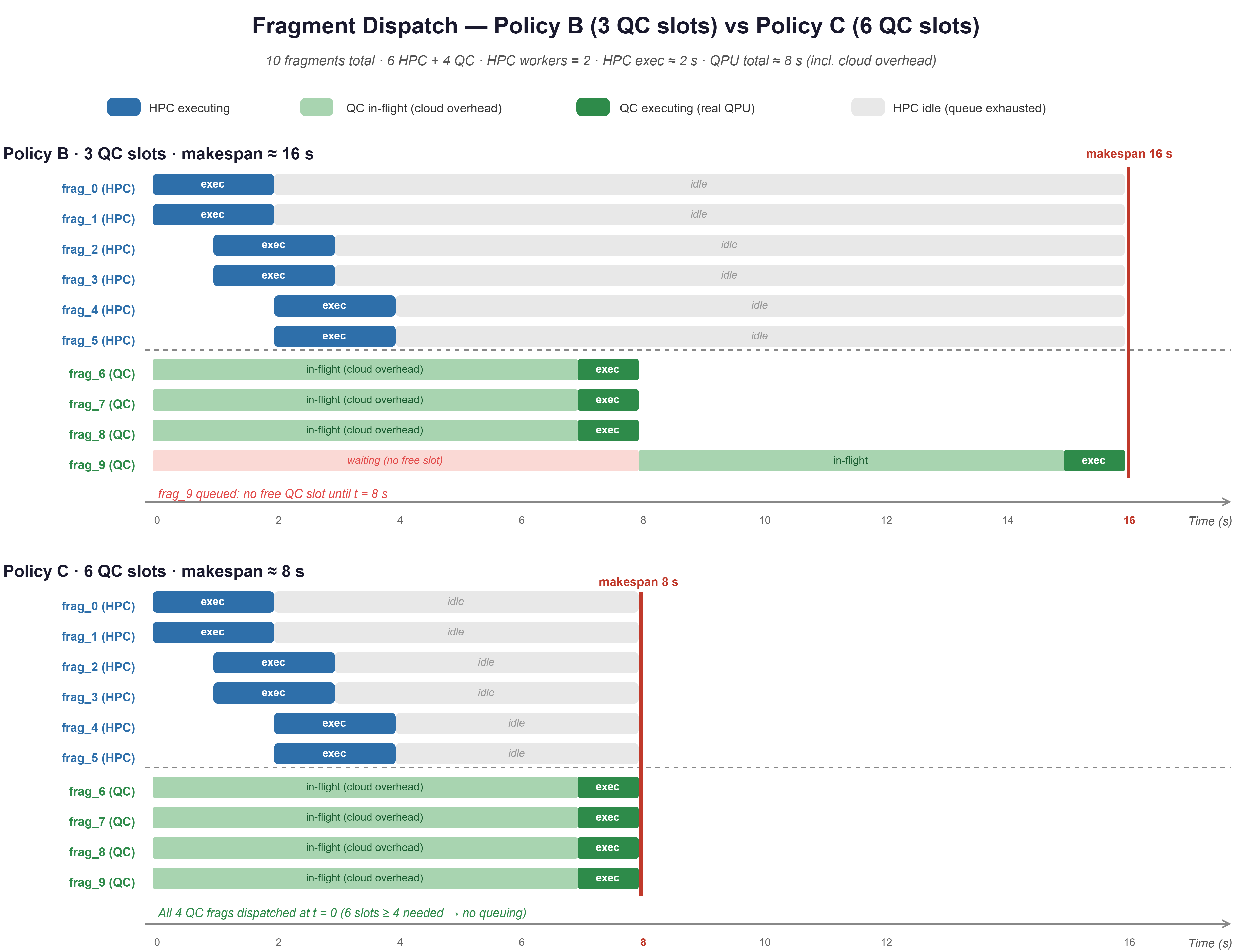}
    \caption{Fragment dispatch comparison between policies B and C.}
    \label{fig:conceptual_frag_dispatch}
\end{figure*}}

\paragraph{Policy~D and diminishing returns.}
Policy~D (30/50/20, 73 MPI ranks, 72 QC slots, 4 nodes) is the most aggressively provisioned configuration. With 73 MPI ranks homogeneously spread across 4 nodes (72 workers plus one coordinator) and 72 QC slots, the planner can commit all 72 fragments in a single dispatch wave. The 22 QC-labeled fragments fit within one QPU round, while the remaining 50 HPC and Undecided fragments are distributed across the pool of workers. The DQR-time of 45.6~s is the shortest among the DQR runs, and the total makespan of 54.9~s improves on the baseline by 2.4\% because almost every rank receives at least one fragment per wave.

Despite this favorable scenario, Policy~D’s DQR-time is 3.9~s longer than Policy~A’s (45.6 vs.\ 41.7~s), even though Policy~D allocates six times as many workers and dispatches all 22 QC fragments in a single wave. The cause is not coordination overhead---$C_{\text{fixed}}$ is comparable across policies (4.4\,s for Policy~A, 5.8\,s for Policy~D)---but the variance of
$t_{\text{round-trip}}$ on IBM Cloud. Under DQR's non-blocking dispatch, each slot is refilled immediately when a fragment completes; there are no synchronization barriers between logical rounds. $T_{\text{QC}}$ is therefore determined by the \emph{slowest} fragment in the entire job, not by the mean round duration. With 22 fragments dispatched concurrently, the standard deviation of $t_{\text{round-trip}}$ rises to 8.4\,s ($\bar{t}_{\text{QPU}} = 26.2$\,s), against 1.2\,s ($\bar{t}_{\text{QPU}} = 13.0$\,s) for Policy~A's 7 fragments; the slowest fragment sets $T_{\text{QC}} = 39.8$\,s, exceeding Policy~A's $T_{\text{QC}} = 37.2$\,s. This is a property of the variance of $t_{\text{round-trip}}$ on IBM Cloud, not of the DQR dispatch architecture.

\begin{figure}
    \centering
    \includegraphics[clip,width=0.999\linewidth,trim={0.2cm 0.2cm 0.2cm 1cm}]{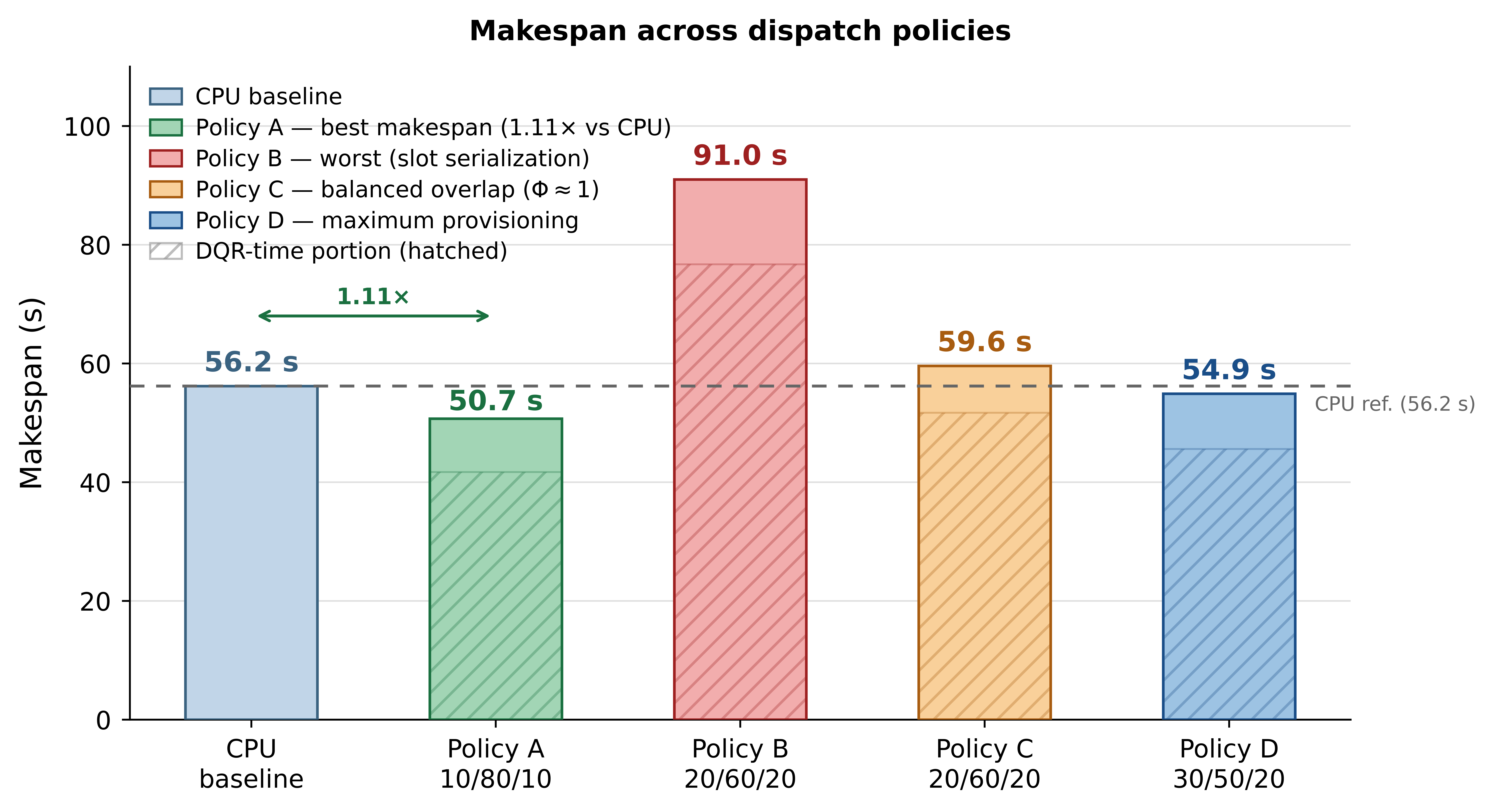}
    \caption{Makespan (s) for the CPU baseline and DQR Policies~A--D on the 32-qubit HEA circuit. Hatching marks the DQR-time portion of each bar, and the dashed line marks the CPU baseline reference (56.2~s).}
    \label{fig:makespan}
\end{figure}

Figure~\ref{fig:makespan} summarizes makespan for all five configurations. The DQR-time reduction from Policy~B to Policy~D (76.7~s $\to$ 45.6~s) is explained by two independent levers: increasing QC slots from 3 to 72 removes slot-induced QC serialization, and scaling MPI processes from 13 to 73 (one worker per fragment plus the coordinator) eliminates rank contention. With only 3 QC slots, fragments queue sequentially for QPU access; with 72 slots, all QC fragments can be concurrently in flight. Likewise, with 12 MPI rank workers managing 72 fragments, ranks must serialize fragment processing; with 72 workers, this bottleneck disappears and only coordination overhead remains.

\review{
The behavior of the four policies is captured by a simple round model of
the QC critical path. With $n_{\mathrm{QC}}$ QC-labeled fragments, $s$ QC
slots and mean round-trip time $\bar{t}_{\mathrm{QPU}}$,
\begin{equation}
  \hat{T}_{\mathrm{QC}} = \left\lceil \frac{n_{\mathrm{QC}}}{s} \right\rceil \bar{t}_{\mathrm{QPU}} .
  \label{eq:tqc_model}
\end{equation}
For Policies~A--C, Eq.~\eqref{eq:tqc_model} predicts $\hat{T}_{\mathrm{QC}} = 39.0$,
$75.0$ and $49.2$~s against observed values of $37.2$, $72.4$ and $47.0$~s,
i.e.\ within 5\%. Since the coordination residual
$C_{\mathrm{fixed}}$ remains between $2.5$ and $5.8$~s across all four policies,
DQR-time follows $\max(\hat{T}_{\mathrm{QC}}, T_{\mathrm{HPC}}) + C_{\mathrm{fixed}}$.
The QC fraction and the slot count can therefore be chosen before job
submission from the expected QPU latency of the target backend.
The model underestimates Policy~D ($26.2$~s predicted versus $39.8$~s
observed): when 22 fragments are dispatched in a single wave, $T_{\mathrm{QC}}$
is set by the slowest fragment rather than by the mean, and the standard
deviation of $t_{\mathrm{round\text{-}trip}}$ on IBM Cloud ($8.4$~s) dominates.
Selecting a policy automatically thus requires accounting for the variance
of the QPU round-trip time, not only for the number of QPU rounds.
}

Across all DQR configurations, $t_{\text{setup}_h}$ ranges from 7.8~s (Policy~C) to 14.3~s (Policy~B), corresponding to 13--18\% of the makespan (Table~\ref{tab:summary}). 
Although ideally, $t_{\text{setup}_h}$ should be constant, in production we observed run--to--run variation, in which the dominant contributor is gRPC server startup (3.1--8.3~s), which includes Python environment activation and Qdislib initialization and varies across runs due to node-local caching and filesystem load. Cutting and labeling together consume under 3.5~s in every run. 
This overhead is independent from DQR, but it has to be accounted for in the makespan of the job.

\subsubsection{Scaling to deeper circuits.}\label{subsubsec:perf_res_sca}
To probe the framework’s behavior at higher depth, we repeat the experiment on a 32-qubit HEA circuit with two layers ($L{=}2$), totaling 192 gates and depth~68. Doubling the number of layers doubles the number of $\mathrm{CZ}$ gates from 32 to 64, increasing the number of cuts from $k{=}2$ to $k{=}4$. Under gate cutting with base~6, this yields $6^4 = 1296$ reconstruction terms and $2\times 6^4 = 2,592$ subcircuit evaluations, a $36\times$ increase in pipeline volume relative to the $L{=}1$ case. The global reconstruction factor becomes $1/2^4 = 0.0625$, consistent with four independent $\mathrm{CZ}$ cuts. Subcircuit generation dominates QCut worker time at 65.8~s out of 76.3~s total, reflecting the $\approx 70\times$ increase in serialization work compared to $L{=}1$. Results for the CPU-only baseline and the DQR run appear in Tables~\ref{tab:l2_comparison} and~\ref{tab:qc_latency_sim}.

The CPU-only baseline completes in 96.4~s, spending 79.8~s on Qulacs \texttt{state vector} simulation and 15.8~s on sampling. The DQR configuration uses a 5/80/15 labeling policy. The 5\% QC budget is a deliberate resource cap that limits QPU consumption to the 130 fragments most suited for quantum execution, given the high $\bar{t}_{\text{QPU}}$ expected from the cloud backend at this fragment volume. The pipeline uses 15 QC slots and 193~MPI ranks (192 workers plus one coordinator) across 3 nodes. Of the $2,592$ fragments, 130 are labeled QC, $2,074$ HPC, and 388 Undecided. The total makespan is 316.7~s, a $3.3\times$ slowdown relative to the CPU baseline, with a DQR-time of 190.4~s and $t_{\text{setup}_h} = 126.3$~s dominated by QCut server startup and subcircuit generation.

\begin{table}
\centering
\caption{CPU-only versus DQR pipeline for the 32-qubit HEA at $L{=}2$ (192~gates, depth~68, 1024~shots).}
\label{tab:l2_comparison}
\setlength{\tabcolsep}{2pt}
\renewcommand{\arraystretch}{1.05}
\resizebox{0.9\columnwidth}{!}{
\begin{tabular}{@{}lrr@{}}
\toprule
\textbf{Metric}
  & \textbf{CPU baseline (L2)}
  & \textbf{DQR (L2)} \\
\midrule
\rowcolor{fgcs@header}
\multicolumn{3}{@{}l}{\textit{Configuration}} \\
Circuit ($n$, gates, depth) & 32 q, 192, 68 & 32 q, 192, 68 \\
QC/HPC/U (\%)       & ---          & 5/80/15 \\
QC slots            & ---          & 15 \\
MPI ranks (\textit{np}) & ---      & 192+1 \\
HPC Nodes           & 1            & 3 \\

\midrule
\rowcolor{fgcs@header}
\multicolumn{3}{@{}l}{\textit{Performance}} \\
CPU simulation (s)  & 79.8         & --- \\
Sampling + EV (s)   & 15.8         & --- \\
Makespan (s)        & 96.4         & 316.7 \\
DQR-time (s)        & ---          & 190.4 \\
$T_{\text{QC}}$ (s) & ---          & 181.3 \\
$T_{\text{HPC}}$ (s)& ---          & 76.7 \\
$\bar{t}_{\text{QPU}}$ (s)
  & --- & $18.5{\pm}3.0$ \\
$\bar{t}_{\text{HPC}}$ (s)
  & --- & $4.6{\pm}2.3$ \\
\midrule
\rowcolor{fgcs@header}
\multicolumn{3}{@{}l}{\textit{Results}} \\
$\langle O\rangle$          & $+0.031$    & $+0.002$ \\
\rowcolor{fgcs@best}
vs CPU-only                   & ---         & $-3.29\times$ \\
\bottomrule
\end{tabular}
}
\end{table}

\begin{figure}
    \centering
    \includegraphics[clip,width=0.70\linewidth,trim={0.2cm 2.27cm 17.8cm 3.2cm}]{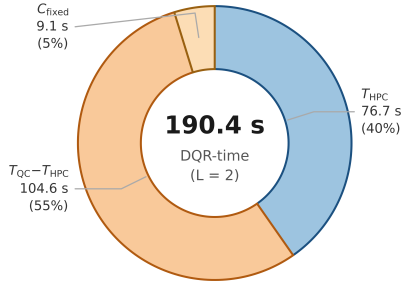}
    \caption{DQR-time decomposition for the $L{=}2$ run: $T_{\text{HPC}} = 76.7$~s (40\,\%) runs in parallel with the 130~QC fragments; the remaining 55\,\% corresponds to $T_{\text{QC}} - T_{\text{HPC}} = 104.6$~s, with $C_{\text{fixed}} = 9.1$~s (5\,\%) for MPI coordination and tensor reconstruction.}
    \label{fig:dqrtime-decomposition}
\end{figure}

The $3.3\times$ slowdown is primarily due to high $\bar{t}_{\text{QPU}}$ on IBM Quantum Cloud, not to the DQR dispatch logic. With 15 QC slots and $n_{\text{QC}}=130$ fragments, the asynchronous dispatcher requires $\lceil 130/15 \rceil = 9$ effective dispatch rounds. Each round completes when the slowest in-flight fragment returns; with $\bar{t}_{\text{QPU}} = 18.5$~s, $T_{\text{QC}} = 181.3$~s constitutes 95\% of DQR-time $= 190.4$~s, with $C_{\text{fixed}} = 9.1$~s (5\%) covering MPI coordination and tensor reconstruction. Figure~\ref{fig:dqrtime-decomposition} decomposes the observed DQR-time into $T_{\text{HPC}}$, $T_{\text{QC}}$, and $C_{\text{fixed}}$.

\begin{figure}
    \centering
    \includegraphics[clip,width=0.9\linewidth,trim={10.5cm 0.2cm 0.2cm 0.2cm}]{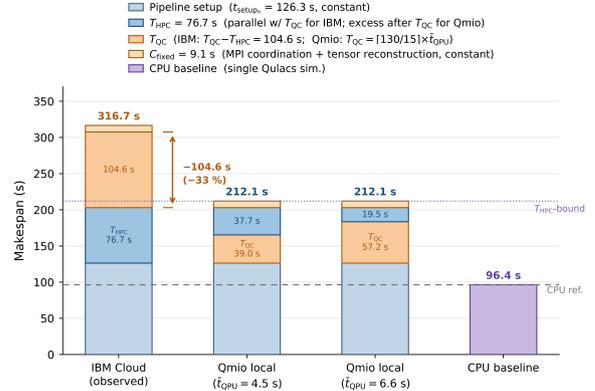}
    \caption{Makespan sensitivity to $\bar{t}_{\text{QPU}}$ for the $L{=}2$ run. Pipeline ($t_{\text{setup}_h}$) and $T_{\text{HPC}}$ remain constant; substituting $\bar{t}_{\text{QPU}} = 18.5$~s (IBM Cloud) with $\bar{t}_{\text{QPU}} = 4.5$ or $6.6$~s (Qmio) reduces makespan by $104.6$~s (33\,\%). } 
    \label{fig:makespan-sensitivity}
\end{figure}

\paragraph{Sensitivity to $\bar{t}_{\text{QPU}}$.}
To assess sensitivity to $\bar{t}_{\text{QPU}}$, we replace the $\bar{t}_{\text{QPU}}$ (IBM Cloud) with the per-fragment times measured on the on-premises CESGA Qmio QPU (Table~\ref{tab:qc_latency_sim}) $\bar{t}_{\text{QPU}} = 6.6$~s (mean) and 4.5~s (median) per fragment, obtained in the local QPU experiment (Table~\ref{tab:local_eval}). 
\review{
These rows are projections rather than measurements: a real $L{=}2$ run on
Qmio is not currently possible, because the native stack rejects the
mid-circuit \texttt{if\_else} branches of gate-cut fragments
(Section~\ref{subsubsec:local_qpu}) and exposes a single serialized QPU slot. The
Qmio rows therefore describe a hypothetical backend combining Qmio's
per-fragment latency with 15 concurrent slots. The round model used for the
projection predicts the QC critical path of Policies~A--C within 5\% and
underestimates the observed IBM run by 8\% (166.5~s versus 181.3~s), so the
projected makespans should be read as moderately optimistic.
}
All other parameters (slot count, MPI allocation, subcircuit volume) remain unchanged. Under the Qmio median $\bar{t}_{\text{QPU}} = 4.5$~s, the QC critical path reduces to $9 \times 4.5 = 40.5$~s, and under the mean (6.6~s) to 59.4~s. In both cases $T_{\text{QC}} < T_{\text{HPC}} = 76.7$~s (HPC-bound), so DQR-time $= T_{\text{HPC}} + C_{\text{fixed}} = 76.7 + 9.1 = 85.8$~s and the total makespan drops to $t_{\text{setup}_h} + 85.8 = 212.1$~s. This corresponds to a saving of 104.6~s (33\% of the observed makespan) $[{=}\,T_{\text{QC}}^{\,\text{IBM}} - T_{\text{HPC}}]$ and a $2.20\times$ factor of the CPU baseline. Figure~\ref{fig:makespan-sensitivity} compares the observed IBM Cloud makespan with the Qmio-local projections and the CPU reference.

Provisioning additional QC slots beyond 15 brings no measurable improvement in this regime: $T_{\text{QC}}$ (40.5 or 59.4~s) already completes well before $T_{\text{HPC}} = 76.7$~s. The system is therefore HPC-bound for any practical slot count, and extra QC concurrency lies off the critical path. Using experimentally observed values, this demonstrates that the DQR dispatch architecture introduces no fundamental latency penalty: the bottlenecks are high $\bar{t}_{\text{QPU}}$ (IBM Cloud) and HPC throughput, not the orchestration layer.

\begin{table}
\centering
\caption{$\bar{t}_{\text{QPU}}$ sensitivity for the $L{=}2$ DQR run.
         IBM row: observed statistics (mean\,$\pm$\,std\,/\,median\,/\,range)
         from the coordinator log.
         Qmio rows: sensitivity projections using single $\bar{t}_{\text{QPU}}$
         values from on-premises measurements
         (see Table~\ref{tab:local_eval}). $C_{\text{fixed}} = 9.1$~s (from the IBM run) is held constant across all Qmio rows.}
\label{tab:qc_latency_sim}
\setlength{\tabcolsep}{4pt}
\renewcommand{\arraystretch}{1.05}
\resizebox{\columnwidth}{!}{%
\begin{tabular}{@{}lrrrrrrr@{}}
\toprule
\textbf{Configuration}
  & \textbf{$\bar{t}_{\text{QPU}}$ (s)}
  & \textbf{$T_{\text{QC}}$ (s)}
  & \textbf{DQR-time (s)}
  & \textbf{Makespan (s)}
  & \textbf{$\Phi$}
  & \textbf{vs CPU} \\
\midrule
\rowcolor{fgcs@header}
\multicolumn{7}{@{}l}{\textit{Observed (IBM Cloud, $n_{\text{QC}}=130$)}} \\
IBM Torino
  & $18.5{\pm}3.0$ / 18.0 / 12.6--28.0
  & 181.3 & 190.4 & 316.7 & 2.36 & ${\times}3.29$ \\
\midrule
\rowcolor{fgcs@header}
\multicolumn{7}{@{}l}{\textit{Qmio on-premises — sensitivity}} \\
$\bar{t}_{\text{QPU}} = 16.0$~s (max)
  & 16.0 & 138.7 & 147.8 & 274.1 & 1.81 & ${\times}2.84$ \\
$\bar{t}_{\text{QPU}} = 6.6$~s (mean)
  & 6.6  &  57.2 &  85.8 & 212.1 & 0.75 & ${\times}2.20$ \\
$\bar{t}_{\text{QPU}} = 4.5$~s (median)
  & 4.5  &  39.0 &  85.8 & 212.1 & 0.51 & ${\times}2.20$ \\
\quad Saving vs.\ IBM
  & \multicolumn{6}{r}{$-104.6$~s (33\,\%) for
    $\bar{t}_{\text{QPU}} \leq 6.6$~s} \\
\midrule
\rowcolor{fgcs@header}
\multicolumn{7}{@{}l}{\textit{Reference}} \\
CPU-only (Qulacs)
  & --- & --- & --- & 96.4 & --- & ref. \\
\bottomrule
\end{tabular}%
}
\end{table}

\paragraph{Mixed-latency behavior.}
The five configurations reported above, taken together, expose three recurring patterns that govern DQR-time in the mixed-latency regime ($\bar{t}_{\text{QPU}} \gg \bar{t}_{\text{HPC}}$). We characterize each pattern using the \emph{DQR balance number} $\Phi = T_{\text{QC}} / T_{\text{HPC}}$, computable directly from the coordinator log as the ratio of the last QC and last HPC fragment completion timestamps, and the \emph{overlap speedup} which satisfies $\sigma \in (1, 2]$: the lower bound is strict because both paths take non-zero time, and $\sigma = 2$ is attained only at $\Phi = 1$, when both paths complete simultaneously. $\sigma$ measures the makespan gain of concurrent versus hypothetical sequential dispatch.

The QC slot count $s$ is the primary tuning lever, not the rank count. When $\Phi > 1$ (QC-bound configurations: Policies~A, B, D and the observed $L{=}2$ run), DQR-time is dominated by $T_{\text{QC}}$: $T_{\text{HPC}}$ has already completed, so adding MPI ranks cannot reduce $T_{\text{QC}}$. Policy~D illustrates the cost of ignoring this: with 72 workers (a $6\times$ increase over Policy~A) and $4\times$ the node-hours, its DQR-time is still 3.9~s longer because MPI coordination overhead outweighs any residual HPC parallelism. The decisive lever is~$s$: halving the number of effective QC rounds by doubling~$s$ (Policy~B $\to$ C) recovers 31.4~s of makespan at identical rank count and node allocation.

Policy~C ($\Phi = 0.96$, $\sigma = 1.96$) is the only $L{=}1$ configuration that is HPC-bound and achieves the best overlap among all four policies---$\sigma$ is closest to~2, and $C_{\text{fixed}} = 2.57$~s is the smallest of the four, reflecting that both critical paths nearly clear simultaneously with minimal coordination residual. Despite this, Policy~C produces a longer DQR-time (51.7~s) than Policy~A (41.7~s). The reason is not coordination overhead but the absolute magnitude of $T_{\text{QC}}$: Policy~C dispatches $n_{\text{QC}} = 29$ fragments at $\bar{t}_{\text{QPU}} = 24.6 \pm 6.0$~s per fragment, requiring $\lceil 29/20 \rceil = 2$ effective QPU rounds and yielding $T_{\text{QC}} = 47.0$~s---longer than Policy~A's $T_{\text{QC}} = 37.2$~s (7~fragments, $\bar{t}_{\text{QPU}} = 13.0$~s, 3~rounds). Maximizing~$\sigma$ (targeting $\Phi \approx 1$) is therefore not equivalent to minimizing makespan: perfect overlap is achievable while the parallel paths themselves are long. The best-makespan configurations (Policy~A, $\sigma = 1.48$; $L{=}2$ observed, $\sigma = 1.42$) operate at $\Phi \approx 2$--$2.4$, safely QC-bound, with $C_{\text{fixed}} < 5$~s and $T_{\text{QC}}$ kept short through a small, low-$\bar{t}_{\text{QPU}}$ QC fragment set.

Finally, the operative bottleneck depends on the $\bar{t}_{\text{QPU}}$ regime, not only on the labeling policy. The $L{=}2$ sensitivity analysis (Table~\ref{tab:qc_latency_sim}) shows that the same DQR configuration (15 slots, 193 ranks) transitions from QC-bound ($\Phi = 2.36$, IBM Cloud) to HPC-bound ($\Phi < 1$, Qmio on-premises) solely due to a change in $\bar{t}_{\text{QPU}}$ from 18.5~s to 4.5--6.6~s. Under high $\bar{t}_{\text{QPU}}$ (IBM Cloud), reducing $\bar{t}_{\text{QPU}}$ is the dominant lever, and additional QC concurrency is largely irrelevant; under low $\bar{t}_{\text{QPU}}$ (Qmio), the bottleneck shifts to HPC throughput, and extra MPI ranks would reduce DQR-time. Provisioning decisions---how many slots and ranks to allocate---must therefore be calibrated to the actual $\bar{t}_{\text{QPU}}$ of the target backend, not merely to circuit size or fragment count.

\section{Discussion}\label{sec:disc}
The experiments confirm five key properties of the DQR architecture, demonstrating its viability for production HPC--QC workloads under realistic NISQ constraints.

First, the non-blocking wave-based dispatcher effectively overlaps $T_{\text{HPC}}$ with $T_{\text{QC}}$. Policy~A achieves a $1.11\times$ makespan speedup over the monolithic CPU baseline while co-executing fragments on the high-latency IBM~Torino ($\bar{t}_{\text{QPU}} \approx 13.0$~s per fragment). As Policy~A is QC-bound ($\Phi = 2.08$), $T_{\text{HPC}} = 17.9$~s completes within the QC critical path---a direct consequence of the pipelined wave model, where new dispatches overlap prior executions without global barriers.

Second, the Policy~B-vs.~C comparison ($20/60/20$ labeling fixed; QC slots $3 \to 20$) isolates slot capacity as the primary tuning lever, slashing makespan by 34.5\% ($91.0 \to 59.6$~s) by collapsing 14 QC fragments from five serialized rounds into two concurrent ones (29 total QC-dispatched fragments). This tunability---independent control of labeling fraction and slot count---enables precise load balancing, absent in static frameworks like Qdislib or early XACC integrations.

Third, the Undecided label serves as an emergent load balancer. Deferred until labeled fragments clear, Undecided fragments opportunistically fill freed HPC/QC slots based on live capacity, mitigating imbalances from variance in $t_{\text{round-trip}}$ or transient degradations ($C_{\text{fixed}} = 2.57$~s for Policy~C, the smallest of the four policies).

Fourth, per-fragment fault isolation ensures resilience. In the local Qmio run, two gate-cut fragments fail with ``\texttt{if\_else} not supported'' errors, exhaust retries, failover to HPC via policy, and complete transparently---preserving the full 72-fragment pipeline without restart or user intervention. This granularity exploits fragment independence, scaling fault tolerance beyond whole-job checkpoints.

Fifth, overhead scales favorably: at $L=2$ ($2,592$ fragments, 193 ranks), $C_{\text{fixed}} = 9.1$~s accounts for only 5\% of DQR-time; the remaining 95\% is $T_{\text{QC}}$ driven by $\bar{t}_{\text{QPU}}$ on IBM Cloud (Figure~\ref{fig:dqrtime-decomposition}). The sensitivity analysis confirms backend swapability: substituting $\bar{t}_{\text{QPU}} = 4.5$--$6.6$~s (Qmio) reduces makespan to 212.1~s ($2.20\times$ the CPU baseline), transitioning from QC-bound ($\Phi=2.36$) to HPC-bound ($\Phi<1$) without architectural changes.

\review{
Sixth, the reported behavior stems from the dispatch abstraction rather than
from the MPI transport. All transport-related cost is contained in
$C_{\mathrm{fixed}}$, which also includes tensor reconstruction and amounts
to 8.7\% (Policy~A), 10.6\% (Policy~D) and 2.9\% ($L{=}2$) of the makespan;
even a zero-overhead runtime could not change the makespan beyond these
fractions.
}

\review{
These properties come with a caveat that the evaluation makes explicit: a
wave-based dispatcher does not by itself guarantee an improvement over the
baseline. With the same coordinator and the same circuit, Policies~A and~D
improve on the CPU reference while Policies~B and~C do not. The QC fraction
and the QC slot count jointly determine the outcome, and they must be
calibrated to the QPU round-trip time of the target backend, including its
variance (Eq.~\eqref{eq:tqc_model}). DQR makes this calibration explicit and
tunable, but does not yet perform it automatically.
}

\review{
The HPC-heavy splits evaluated here reflect the QPU quota and cloud latency
available to this study rather than a design constraint. In a QPU-dominant
regime, a QC-dominant Autobudget triplet (e.g.\ $(0.9, 0, 0.1)$) reduces the
HPC pool to Undecided overflow and failover targets, and DQR-time is governed
by Eq.~\eqref{eq:tqc_model} with $\Phi \gg 1$: performance depends on the QC
slot count and QPU latency, and the HPC allocation need only cover the
expected failovers, approximately $p_{\mathrm{fail}}\, n_{\mathrm{QC}}$. This
regime is most relevant when the full circuit exceeds QPU capacity; there,
fragments may also exceed classical simulation limits, in which case the
admissible-backend set $\mathcal{B}_i = \{\mathrm{QC}\}$ prevents relabeling
to HPC and restricts recovery to QPU retries.
}

Three concrete limitations emerge from the evaluation. First, a Qmio Cluster compiler gap: Qdislib's gate-cut decomposition injects mid-circuit classical \texttt{if\_else} branches that the current Qmio native stack does not support, preventing fully local hybrid execution and forcing all QC-labeled fragments in the on-premises scenario to fail over to HPC. Resolving this requires either a control-flow-aware native compiler for Qmio, a library update that enables simultaneous compatibility between Qiskit Runtime and Qmio, or a cutting strategy that avoids mid-circuit branches. Second, a provisioning gap: slot count and MPI ranks are fixed at job submission, preventing runtime adaptation to observed $\bar{t}_{\text{QPU}}$ variance; Policy~D demonstrates that over-provisioning ranks increases $C_{\text{fixed}}$ without reducing $T_{\text{QC}}$ when the system is QC-bound.
\review{
Third, a benchmark-coverage gap. Gate-cutting the regular HEA circuit yields structurally homogeneous fragments, so the evaluation relies on the Autobudget policy, and the metric-driven labeling policies (Budget, Score and Hybrid) are not exercised on real workloads. Workloads with heterogeneous fragments, such as QAOA on irregular graphs or UCCSD-type chemistry ansätze, would exercise them. At 32 qubits, however, these workloads require more gate cuts, and the subcircuit count grows as $6^k$: our L=2 run (k=4) already produced 2,592 evaluations, and k=6 would produce 93,312. This places such a campaign beyond the QPU allocation available for this study.}

\section{Conclusions}\label{sec:conc}
We presented DQR, a runtime framework that decouples quantum circuit cutting from HPC execution orchestration by treating cut fragments as independent schedulable units described by backend-agnostic descriptors. This transforms quantum execution into a classical heterogeneous scheduling problem addressable with mature HPC techniques.
 
Experiments on the CESGA Qmio cluster confirm three quantitative properties. First, the non-blocking wave-based dispatcher overlaps $T_{\text{HPC}}$ with $T_{\text{QC}}$: Policy~A achieves a $1.11\times$ makespan improvement over a monolithic CPU baseline while co-executing fragments on IBM~Torino ($\bar{t}_{\text{QPU}} \approx 13.0$~s, $\Phi = 2.08$). Second, per-fragment fault recovery is transparent: two QPU rejections due to unsupported control-flow are detected, failed over to HPC, and resolved without pipeline restart, demonstrating correct fragment-level isolation. Third, coordination overhead remains bounded: at $2,592$ fragments and 193~MPI ranks ($L{=}2$), $C_{\text{fixed}} = 9.1$~s accounts for only 5\% of DQR-time; the remaining 95\% is $T_{\text{QC}}$ driven by $\bar{t}_{\text{QPU}}$ on IBM Cloud---an infrastructure constraint external to the framework.

These results directly address HPC--QC gaps: (i) cutting-execution decoupling via agnostic descriptors, enabling Qdislib evolution without re-orchestration; (ii) dynamic re-planning for unreliable QPUs, beyond static DAGs; (iii) production primitives (Slurm integration, MPI scalability) absent in research prototypes.

\review{
The limitations identified in Section~\ref{sec:disc} define the next steps of this work. First, the provisioning gap will be addressed by integrating DQR with the DMR malleable resource manager ~\cite{iserte_resource_2025}, so that MPI ranks and QC slots can be resized at runtime as the observed QPU round-trip time evolves; the multi-backend QPU path needed to dispatch to local and remote QPUs simultaneously is already implemented. Second, the compiler gap requires either a control-flow-aware native compiler for Qmio or a cutting strategy that avoids mid-circuit classical branches, enabling fully on-premises hybrid execution. 
Third, the labeling policies will be evaluated on workloads that produce heterogeneous fragments, such as QAOA on irregular graphs and chemistry ansatz, where the metric-driven policies (Budget, Score, Hybrid) become decisive.
Fourth, the calibration model of Eq.~\eqref{eq:tqc_model} predicts the QC critical path from the QC fraction, the slot count and the mean QPU latency, but underestimates it when many fragments share a single wave; extending it with the variance of the
round-trip time would enable automatic policy selection before job
submission, and learning-based routing could further adapt the policy at
runtime from observed completions. 
Fifth, porting the dispatch layer to a task-based runtime would directly
quantify the contribution of the transport layer.
Finally, a repeated end--to--end campaign across on-premises and cloud backends would complement the latency-substitution analysis of Section~\ref{subsubsec:perf_res_sca} with direct measurements.
}
 
\section*{Acknowledgements}
\review{The authors thank the anonymous reviewers whose suggestions significantly improved the quality of this manuscript,} 
and Accelcom@BSC for the provided support.

Language polishing was performed using an AI language model, with subsequent thorough review and editing by the authors. The authors take full responsibility for the final manuscript content.

\section*{CRediT authorship contribution statement}
\begin{itemize}
    \item Ricard S. Raigada-García: Writing - Original Draft, Software, Methodology, Investigation, Formal analysis.
    \item Josep Jorba: Writing - Review \& Editing, Project Administration.
    \item Sergio Iserte: Writing - Review \& Editing, Resources, Investigation, Conceptualization.
\end{itemize}



\section*{Funding sources}

This research was made possible through the access granted by the Galician Supercomputing Center (CESGA) to its Qmio quantum computing infrastructure with funding from the European Union, through the Operational Programme Galicia 2014-2020 of ERDF\_REACT EU, as part of the European Union’s response to the COVID-19 pandemic.

\section*{Data availability}
The framework developed for this work, DQR, is available \href{https://github.com/ToroData/Dynamic-Queue-Router}{https://github.com/ToroData/Dynamic-Queue-Router}

\bibliographystyle{elsarticle-num}
\bibliography{bib/sota}

\end{document}